\documentclass[11pt]{elsarticle}
\biboptions{sort&compress,numbers}
\usepackage{microtype}
\usepackage[margin=1in]{geometry}
\usepackage{amsmath,amssymb,amsthm,bm,stmaryrd}
\usepackage{subcaption}
\biboptions{sort&compress}
\usepackage{upgreek}
\usepackage{epstopdf}
\usepackage{float}
\usepackage{xcolor}
\usepackage[labelfont=bf]{caption}
\usepackage[export]{adjustbox}
\AppendGraphicsExtensions{.tif}

\usepackage[final]{changes} %% UN-COMMENT THIS LINE TO HIDE CHANGES
\definecolor{C0}{HTML}{1F77B4}
\definecolor{C1}{HTML}{FF7F0E}
\definecolor{C2}{HTML}{2ca02c}
\definecolor{C3}{HTML}{d62728}
\definecolor{C4}{HTML}{9467bd}
\definecolor{C5}{HTML}{8c564b}
\definechangesauthor[color=C0]{R1} % changes relevant to R1
\definechangesauthor[color=C1]{R2} % changes relevant to R2
\definechangesauthor[color=C3]{RA} % changes relevant to R1 & R2

\begin{document}

    %==================================================
    % Instructions
    % Use id=R1 for reviewer 1, R2 for reviewer 2, and RA for R1 & R2
    % Indicate query number 1,2,or 3 in comment part.
    % To add text, please use \added[id=R1,comment={}]{<text>}
    % To delete text, please use \deleted[id=R1,comment={}]{<text>}
    % To replace text, please use \replaced[id=R1,comment={}]{<new text>}{<old text>}
    %==================================================

    \begin{frontmatter}

        \title{Transmitting multiple high-frequency phonons across length scales using the concurrent atomistic-continuum method}
        \author[auburn]{Alexander S. Davis}
        \author[auburn]{Vinamra Agrawal\corref{cor1}}
        \ead{vinagr@auburn.edu}
        \cortext[cor1]{Corresponding author}
        \address[auburn]{Department of Aerospace Engineering, Auburn University, Auburn, AL USA}
        
        \begin{abstract}
            Coupled atomistic-continuum methods can describe large domains and model dynamic material behavior for a much lower computational cost than traditional atomistic techniques.
            However, these multiscale frameworks suffer from wave reflections at the atomistic-continuum interfaces due to the numerical discrepancy between the fine-scaled and coarse-scaled models.
            Such reflections are non-physical and may lead to unfavorable outcomes such as artificial heating in the atomistic region.
            In this work, we develop a technique to allow the full spectrum of phonons to be incorporated into the coarse-scaled regions of a periodic concurrent atomistic-continuum (CAC) framework.
            This scheme tracks phonons generated at various time steps and thus allows multiple high-frequency wave packets to travel between the atomistic and continuum regions.
            Simulations performed with this method demonstrate the ability of the technique to preserve the coherency of waves with a range of wavevectors as they propagate through the domain.
            This work has applications for systems with defined boundary conditions and may be extended to more complex problems involving waves randomly nucleated from an impact within a multiscale framework.
        \end{abstract}
    
    \end{frontmatter}
    
    \section{Introduction} \label{Sec: Introduction}
        Multiscale modeling techniques endeavor to link observable material behavior to effects at lower length scales.
        To this end, coupled atomistic-continuum (A-C) frameworks have been developed since the early 1990s to integrate the microscale and macroscale into a single computational domain \cite{kohlhoff1991crack}.
        In particular, \textit{concurrent} A-C methods connect the spatial scales directly such that the continuum region surrounds an inner atomistic region containing the phenomena of interest.
        Some examples of concurrent frameworks include the Coupling of Length Scales (CLS) method \cite{xiao2004bridging}, the Coupled Atomistic Discrete Dislocation (CADD) method \cite{Shilkrot2002Coupled}, and the Quasicontinuum (QC) method \cite{tadmor1996quasicontinuum}.
        One of the central challenges with concurrent schemes is ensuring compatibility at the A-C interfaces so as to mitigate ghost forces in static systems and spurious wave behavior in dynamic systems \cite{xu2018modeling}.
        Typically, such non-physical phenomena arise because the spectrum of the continuum model has a much smaller cutoff frequency than that of the atomistic model \cite{tadmor2011modeling}. 
        Although many techniques have been developed to reduce ghost forces in static frameworks \cite{eidel2009variational,xu2015quasistatic}, the advancement of dynamic multiscale methods is nevertheless hindered by spurious wave reflections at the A-C interfaces. 
        
        To overcome this obstacle, most concurrent methods incorporate techniques to either minimize or absorb transient waves impinging on the A-C interfaces \cite{qu2005finite,karpov2005green,li2007variational,jolley2009modelling}. 
        An early scheme developed by \cite{cai2000minimizing} incorporates Langevin dynamics into the fine-scaled equations of motion and dampens specified particles in a ``stadium" region around the inner atomistic core.
        Specifically, the method couples a one-dimensional atomistic domain to a linear elastic continuum and reduces wave reflections at the A-C interfaces by calculating the time-history-kernel (THK).
        This approach has proven to be effective, and variations of it have been introduced into other concurrent multiscale frameworks such as CADD \cite{qu2005finite} and the Bridging Scale Method (BSM) \cite{wagner2003coupling}.
        However, because the THK method suffers from issues related to computational expense and scalability, various BSM frameworks have developed more efficient THK techniques, but such schemes are still only effective for linear solids \cite{karpov2005green,park2005non,karpov2007phonon}.
        Other approaches to reduce wave reflections include minimizing the reflection coefficient at the A-C boundaries \cite{weinan2001matching,li2007variational} as well as applying digital filters to remove high-frequency phonons that travel back into the fine-scaled region \cite{mathew2011concurrent,ramisetti2013spatial}.
        
        Because all of these methods either minimize or absorb waves impinging on the A-C interfaces, information from short-wavelength phonons is lost.
        Furthermore, damping methods will inevitably eliminate fine-scaled wave data which should instead be transmitted across the boundaries \cite{chen2018passing}.
        One of the first attempts to solve this problem came in \cite{chirputkar2008coupled} which enhances a space-time discontinuous Galerkin finite element method by incorporating an enrichment function into the system.
        This technique has since been used to study both wave and crack propagation through materials, and it can successfully conserve energy and transmit high-frequency waves across the A-C interfaces \cite{yang2012generalized}.
        Unfortunately, the framework in \cite{chirputkar2008coupled} requires extra degrees of freedom in the coarse-scaled regions, and the enriched functions must be removed at the continuum nodes in order to incorporate the short-wavelength phonons.
        Therefore, conserving the correct wave phase is challenging, so this technique cannot be easily used to study dynamic problems which require phonon coherency.
        As a result, a concurrent multiscale method is needed which would preserve phonon coherency and permit the full range of phonons to travel across the A-C interfaces. 
        
        Previous work has developed a technique to transfer high-frequency phonons across length scales within a concurrent atomistic-continuum (CAC) framework \cite{chen2018passing}.
        CAC is a dynamic multiscale method which follows the solid state physics model of crystals whereby the structure is continuous at the lattice level but discrete at the atomic level, and a single set of governing equations is used throughout the entire domain \cite{chen2019concurrent}. 
        As a result, the wave transfer problem reduces to a numerical problem caused by the discrepancy in finite element mesh sizes between the atomistic and continuum regions.
        This is a long-standing obstacle in continuum modeling and was regarded by Zienkiewicz as one of the great unsolved problems in the Finite Element Method \cite{zienkiewicz2000achievements}.
        The work in \cite{chen2018passing} developed a supplemental basis for the CAC solution along with a new lattice dynamics (LD)-based finite element scheme to pass a single high-frequency phonon between the atomistic and continuum regions.
        This technique allowed a wave packet with any wavevector and frequency to travel across the A-C interfaces without introducing new degrees of freedom into the coarse-scaled regions.
        However, this method could only be used for a single phonon and was demonstrated in a non-periodic domain.
        
        In the present article, we develop a technique based upon the work in \cite{chen2018passing} to pass multiple high-frequency phonon wave packets between the atomistic and continuum regions of a periodic CAC framework.
        This method uses the LD interpolation scheme to incorporate short-wavelength displacements into the continuum regions and introduces novel numerical techniques into the formulation to track a variety of wave packets across time.
        Specifically, two Fourier transforms are performed (both before and after the phonon is generated), and the difference in amplitude coefficients are stored in a master array in order to track waves of any wavevector at various time steps. 
        Such a technique will be useful in real-world applications which involve the interaction and transmission of multiple waves within a single atomistic-continuum domain.
        The remainder of this paper is organized as follows: Sec. \ref{Sec: The CAC method} summarizes the finite element implementation of the CAC method; Sec. \ref{Sec: Computational setup} describes the one-dimensional monatomic framework; Sec. \ref{Sec: Wave reflections at the A-C interface} presents simulations performed without the LD formulation and showcases the numerical discrepancy at the A-C interfaces; Sec. \ref{Sec: Lattice dynamics-based finite element formulation} provides a mathematical background of the technique formulated in \cite{chen2018passing} and demonstrates this technique with a single phonon; Sec. \ref{Sec: Upgrading the lattice dynamics technique for multiple waves} gives a detailed explanation of the LD method for multiple waves; Sec. \ref{Sec: Benchmark examples with multiple waves} presents benchmark simulations with multiple phonons within a periodic CAC domain; finally, Sec. \ref{Sec: Conclusion} concludes the article and provides suggestions for future work.
    
    \section{The CAC method} \label{Sec: The CAC method}
        In this section, we discuss the finite element implementation of CAC, and more details can be found in \cite{xiong2009multiscale,deng2010coarse,xiong2011coarse}.
        The mathematical foundation of CAC is Atomistic Field Theory (AFT), and the governing equations of AFT are \added[id=R1,comment={2}]{ensemble averages of} partial differential equations which are similar in form to the balance laws of classical continuum mechanics \added[id=R1,comment={2}]{\cite{chen2005atomistic,chen2009reformulation}}. 
        \added[id=R1,comment={2}]{Recent work has reformulated these equations using the mathematical theory of distributions in which the quantity definitions as well as the balance equations themselves are valid instantaneously \cite{chen2019concurrent}.}
        As in continuum mechanics, the analytical solution to these equations is not readily obtainable, and thus we utilize numerical schemes such as the finite element method (FEM) to solve them. 
        \added[id=R1,comment={2}]{In this work, we refer to such a formulation as `the CAC method'.}
        
        Using the standard definitions of internal force density and kinetic temperature as derived in \cite{chen2005nanoscale} and \cite{chen2006local}, we can rewrite the instantaneous balance equation of linear momentum as follows \cite{xiong2009multiscale}:
        \begin{equation} \label{Eq: FEM Governing Equation 2} 
            \rho^{\alpha} \Ddot{\textbf{u}}^{\alpha}(\textbf{x}) = \textbf{f}_{int}^{\alpha}(\textbf{x}) + \textbf{f}^{\alpha}(\textbf{x})
        \end{equation}
        where $\textbf{u}^{\alpha}(\textbf{x})$ is the displacement of the $\alpha^{th}$ atom in the unit cell \added[id=R1,comment={1}]{located at point \textbf{x}}, $\rho^{\alpha} = m^\alpha/\Delta V $ is the volumetric mass density, $m^{\alpha}$ is the mass of the $\alpha^{th}$ atom, $\Delta V$ is the volume of the unit cell, $\textbf{f}_{int}^{\alpha}(\textbf{x})$ is the internal force density, and $\textbf{f}^{\alpha}(\textbf{x})$ is the force density due to external forces and temperature.
        The terms on the right hand side of Eq. (\ref{Eq: FEM Governing Equation 2}) are given by
        \begin{equation} 
            \textbf{f}_{int}^{\alpha}(\textbf{x}) = \int_{\Omega(\textbf{x}')} \sum_{\beta = 1}^{N_a} \textbf{f} \left[\textbf{u}^{\alpha}(\textbf{x}) - \textbf{u}^{\beta}(\textbf{x}') \right] d\textbf{x}', \quad \textbf{f}^{\alpha}(\textbf{x}) = \textbf{f}_{ext}^{\alpha}(\textbf{x}) - \frac{m^{\alpha} k_B}{M \Delta V} \nabla_{\textbf{x}} T^{\alpha}
            \label{Eq: Internal Force Density}
        \end{equation}
        where $\textbf{f}_{ext}^{\alpha}(\textbf{x})$ is the external force density, $M$ is the total mass of the atoms within a unit cell, $T^{\alpha}$ is the kinetic temperature, and $k_B$ is the Boltzmann constant.
        Here, the internal force density is a nonlinear, nonlocal function of relative displacements between neighboring atoms within a given cutoff radius, and it can be obtained exclusively from the interatomic potential function \cite{yang2014concurrent}. 
        
        We calculate the numerical solution of the governing equation (Eq. \ref{Eq: FEM Governing Equation 2}) by discretizing the material with finite elements such that each element contains a collection of primitive unit cells. 
        Furthermore, each finite element node represents a unit cell which is itself populated by a group of atoms. 
        At the lattice level, we use interpolation within an element to approximate the displacement field as follows \cite{xiong2011coarse}:
        \begin{equation} \label{Eq: Approximate Displacement Field}
            \Hat{\textbf{u}}^{\alpha}(\textbf{x}) = \boldsymbol{\Phi}_{\xi}(\textbf{x}) \textbf{U}_{\xi}^{\alpha}.
        \end{equation}
        Here, $\Hat{\textbf{u}}^{\alpha}(\textbf{x})$ is the displacement field for the $\alpha^{th}$ atom within a given element, $\boldsymbol{\Phi}_{\xi}(\textbf{x})$ is the shape function, and $\textbf{U}_{\xi}^{\alpha}$ is the displacement of the $\alpha^{th}$ atom within the $\xi^{th}$ element node. 
        We let $\xi = 1, 2, ..., n$ where $n$ is the total number of nodes in the element.
            
        Using the method of weighted residuals, we obtain the weak form of the governing equation by multiplying Eq. (\ref{Eq: FEM Governing Equation 2}) with a weight function $\boldsymbol{\Phi}_{\eta}(\textbf{x})$ and integrating over the entire domain:
        \begin{equation} \label{Eq: Weak Form of Governing Equation 1}
            \int_{\Omega(\textbf{x})} \left[\rho^{\alpha} \boldsymbol{\Phi}_{\eta}(\textbf{x})  \Ddot{\textbf{u}}^{\alpha}(\textbf{x}) \right] d\textbf{x} = \int_{\Omega(\textbf{x})} \left[ \boldsymbol{\Phi}_{\eta}(\textbf{x}) \textbf{f}_{int}^{\alpha}(\textbf{x}) \right] d\textbf{x} + \int_{\Omega(\textbf{x})} \left[ \boldsymbol{\Phi}_{\eta}(\textbf{x}) \textbf{f}^{\alpha}(\textbf{x}) \right] d\textbf{x}.
        \end{equation}
        Substituting Eqs. (\ref{Eq: Internal Force Density}) and (\ref{Eq: Approximate Displacement Field}) into Eq. (\ref{Eq: Weak Form of Governing Equation 1}), we get 
        the weak form of the governing equation which can be represented in matrix form as
        \begin{equation} \label{Eq: Matrix Form of Governing Equation}
            \textbf{M}^{\alpha} \Ddot{\textbf{U}}_{\xi}^{\alpha} = \textbf{F}_{int}^{\alpha} + \textbf{F}^{\alpha}
        \end{equation}
        where
        \begin{align}
            \textbf{M}^{\alpha} &= \int_{\Omega(\textbf{x})} \left[\rho^{\alpha} \boldsymbol{\Phi}_{\eta}(\textbf{x}) \boldsymbol{\Phi}_{\xi}(\textbf{x}) \right] d\textbf{x} \\
            \textbf{F}_{int}^{\alpha} &= \int_{\Omega(\textbf{x})} \boldsymbol{\Phi}_{\eta}(\textbf{x}) \int_{\Omega(\textbf{x}')} \sum_{\beta = 1}^{N_a} \textbf{f} \left[\boldsymbol{\Phi}_{\xi}(\textbf{x}) \textbf{U}_{\xi}^{\alpha} - \boldsymbol{\Phi}_{\xi}(\textbf{x}') \textbf{U}_{\xi}^{\beta} \right] d\textbf{x}' d\textbf{x} \\
            \textbf{F}^{\alpha} &= \int_{\Omega(\textbf{x})} \left[ \boldsymbol{\Phi}_{\eta}(\textbf{x}) \textbf{f}^{\alpha}(\textbf{x}) \right] d\textbf{x}
        \end{align}
            
        In the present formulation, we approximate the inertial term using the lumped mass matrix.
        Additionally, no external forces are applied and temperature is incorporated through the use of a thermostat as in \cite{xiong2014prediction} and \cite{chen2018passing}. 
        The internal force density $\textbf{F}_{int}^{\alpha}$ is the most computationally demanding term, and we evaluate it numerically using numerical integration. 
        Finally, the second order differential equation (Eq. \ref{Eq: Matrix Form of Governing Equation}) is solved through the velocity Verlet integration algorithm.
        By using this finite element implementation of AFT, a majority of the degrees of freedom in the continuum regions are eliminated. 
        For critical regions where atomistic behavior is important, the finest mesh is used such that the element length is equal to the atomic equilibrium spacing.
        In this way, CAC uses AFT to produce a unified theoretical framework between the atomistic and continuum regions.
        CAC frameworks are defined as AFT domains which contain both fine-scaled and coarse-scaled regions \cite{davis2022moving}.
            
    \section{Computational setup} \label{Sec: Computational setup}
    
        \subsection{Domain geometry} \label{Sec: Domain geometry}
\begin{figure}[htpb]
                \centering
                \includegraphics[width=0.9\textwidth]{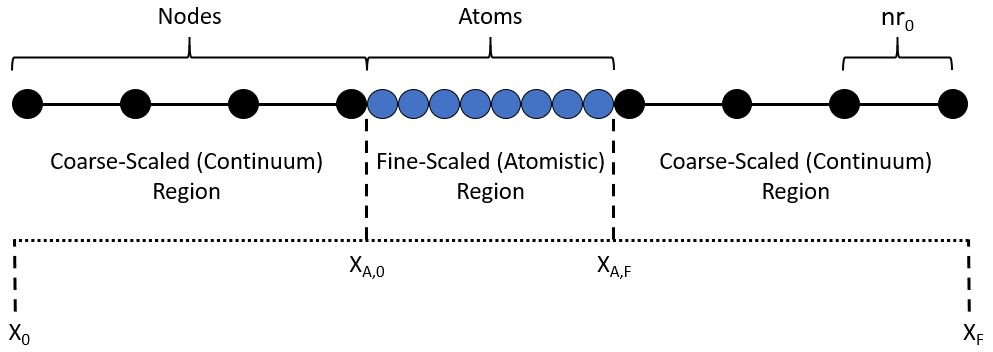}
                \caption{Schematic of the CAC domain \cite{davis2022moving}.}
                \label{Fig:CACSystem}
            \end{figure}

            In this work, the CAC framework uses the conventional finite element formulation with linear interpolation functions discussed above.
            To readily demonstrate the nature of wave transmission and reflection at the A-C interfaces, we develop a one-dimensional CAC domain using an in-house C++ code.
            The monatomic chain consists of $N$ particles which are split into three regions as seen in Fig. \ref{Fig:CACSystem}. 
            The particles in each coarse-scaled (continuum) region are separated by a distance of $nr_0$ and are referred to as \textit{nodes} in the present work. 
            Here, $n$ is some positive integer (6 in this work), and $r_0$ is the equilibrium spacing determined by the potential function. 
            These two coarse-scaled regions flank the inner fine-scaled (atomistic) region on either side.
            The particles in the fine-scaled region are separated by a distance of $r_0$ and are referred to as \textit{atoms} in the present work.
            Because CAC produces a unified atomistic-continuum framework using a single set of governing equations, the atoms and nodes have identical properties with the only difference being their inter-particle spacing. 
            \added[id=R2,comment={1}]{Hence, all force calculations are fully nonlocal, and the interatomic potential is the only constitutive relation \cite{xu2016mesh}.}
            As a result, the particles at the atomistic-continuum interfaces ($x_{A,0}$ and $x_{A,F}$) interact with each other directly \added[id=R2,comment={1}]{without generating ghost forces \cite{xu2015quasistatic,xu2018pycac}}.
            We employ standard periodic boundary conditions in every simulation.
        
        \subsection{Integration algorithm} \label{Sec: Integration algorithm}
            The CAC governing equation (Eq. \ref{Eq: FEM Governing Equation 2}) is a second order ordinary differential equation in time, and we solve it using the velocity Verlet algorithm.
            The time step used in the integration algorithm is chosen to be $\Delta t = 0.001$ ps in order to minimize numerical error.
            
        \subsection{Interatomic potential and material parameters} \label{Sec: Interatomic potential and material parameters}
            We use the modified Morse interatomic potential function to calculate the integrand of the internal force density (Eq. \ref{Eq: Internal Force Density}).
            The standard Morse potential was modified by \cite{macdonald1981thermodynamic} to improve the agreement with experimental values for the thermal expansion of materials.
            The modified Morse potential only considers first nearest neighbor interactions and is given by the following expression \cite{macdonald1981thermodynamic}:
            \begin{equation} \label{Eq: Morse}
                \Pi(r_{ij}) = \frac{D_0}{2B-1}\left[e^{-2 \alpha \sqrt{B} (r_{ij} - r_0)} - 2Be^{-\alpha (r_{ij} - r_0) / \sqrt{B}}\right]
            \end{equation}
            where $r_{ij}$ is the magnitude of the displacement between particle $i$ and $j$, and $r_0$ is the distance at which the potential reaches the minimum.
            We perform simulations with Cu, and the parameters for this material are given in Table \ref{Table: MorsePotentialParameters}.
            Here, we note that $r_0$ is equivalent to the equilibrium spacing along the [110] lattice direction of Cu. 
            \begin{table}[h]
                \centering
                \caption{Material constants and Morse parameters of Cu \cite{macdonald1981thermodynamic}.}
                \label{Table: MorsePotentialParameters}
                \begin{tabular}{||c  c  c  c  c  c  c||}
                \hline
                \textit{Element} & \textit{mass (u)} & \textit{$\rho_0$ (g/$cm^3$)} & \textit{$r_0$ (\AA)} & \textit{$\alpha$ (\AA$^{-1}$)} & \textit{$D_0$ (eV)} & \textit{B} \\
                \hline
                Cu & 63.55 & 8.96 & 2.5471 & 1.1857 & 0.5869 & 2.265 \\
                \hline 
                \end{tabular}
            \end{table}
    
    \section{Numerical discrepancy at the A-C interface} \label{Sec: Wave reflections at the A-C interface}
        In this section, we showcase the numerical discrepancy between the fine-scaled and coarse-scaled regions when modeling high-frequency phonons in a standard CAC formulation.
        To do this, we reproduce the dispersion relation using the domain in Fig. \ref{Fig:CACSystem}.
        
\begin{figure}[htpb]
            \centering
            \includegraphics[width=0.55\textwidth]{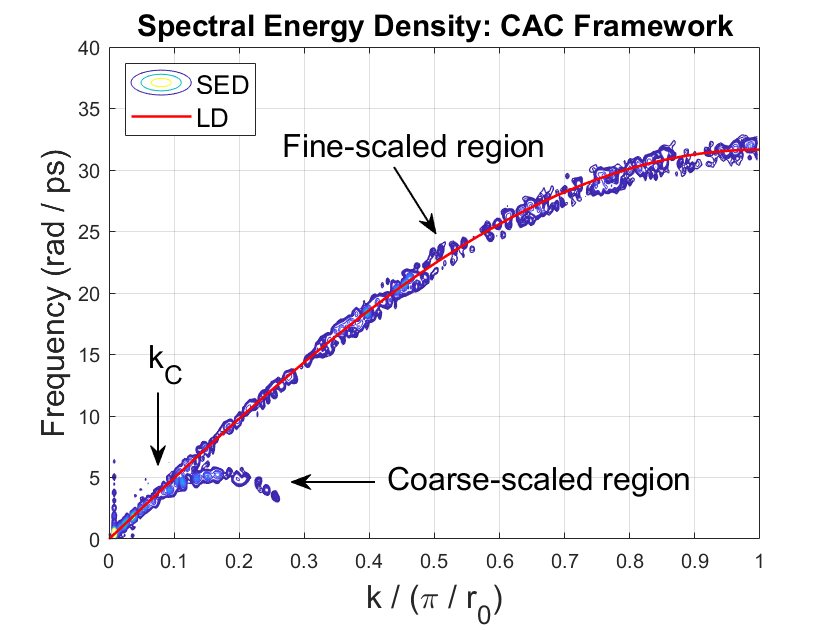}
            \caption{Phonon spectral energy density contour plot of a CAC monatomic chain  calculated using the \textit{Nose-Hoover} thermostat.
            The red line represents the analytical dispersion relation obtained from Lattice Dynamics, and the simulation was performed at $10$ K.}
            \label{Fig:DispersionRelationNH}
        \end{figure}

        The dispersion relation of the CAC framework is obtained by calculating the phonon spectral energy density  which is defined as the average kinetic energy per unit cell as a function of wavevector $k$ and angular frequency $\omega$.
        In 1D, the spectral energy density is given as follows \cite{thomas2010predicting}:
        \begin{equation} \label{Eq: Spectral Energy Density 1D}
            \varphi(k,\omega) = \frac{m}{4 \pi \tau_0 N} \left| \int_0^{\tau_0} \left\{ \sum_{n=1}^{N} \Dot{u}_n(t) \times \mathrm{exp} \left[i k \cdot x_{n}(t_0) - i \omega t \right] \right\} dt \right|^2
        \end{equation} 
        where $\tau_0$ is the total simulation time, $N$ is the total number of particles, $\Dot{u}_n(t)$ is the velocity of particle $n$ at time $t$, and $x_{n}(t_0)$ is the initial position of particle $n$.
        For this simulation, the monatomic chain contains 260 atoms in the fine-scaled region and 20 nodes in each coarse-scaled region for a total of 300 particles, and the domain is maintained at 10 K using the Nose-Hoover thermostat \cite{evans1985nose}.
        Spectral energy density calculations are compared to the analytical dispersion relation obtained from Lattice Dynamics (LD), and this relation for a one-dimensional monatomic crystal is given by the following equation:
        \begin{equation}
            \omega = \sqrt{\frac{4 C}{m}} \left|\mathrm{sin} \left(\frac{k r_0}{2} \right) \right|
        \end{equation}
        where $C$ is the elastic constant defined as the second derivative of the interatomic potential function at $r_0$ in 1D.
        Results are shown in Fig. \ref{Fig:DispersionRelationNH}.
        Here, the contours indicate the magnitude of the spectral energy density for each ($k$, $\omega$) combination, and the red line represents the analytical relation.
        
        In Fig. \ref{Fig:DispersionRelationNH}, we observe that the phonon dispersion relation obtained in the fine-scaled region of the CAC framework is identical to the analytical curve from LD. 
        However, the dispersion relation for the coarse-scaled regions is only accurate for phonons whose wavevector is smaller than a critical value $k_C$.
        This critical wavevector is given by the following equation \cite{chen2017ballistic}:
        \begin{equation}
            k_C = \underset{k}{\operatorname{max}} \left\{ \left| \mathrm{sin} \left(\frac{kr_0}{2} \right) - \mathrm{sin} \left(\frac{kL}{2} \right) \right| \le \epsilon \right\}.
        \end{equation}
        Here, $\epsilon$ is the allowable error, and $L = 6r_0$ is the element length in the coarse-scaled regions.
        We choose an allowable error of $\epsilon = 5$\% which corresponds to a critical wavevector of $k_C = 0.064 \, \pi/r_0$, and a critical wavelength $\lambda_C = 2\pi/k_C = 7.96$ nm.
        Therefore, only phonons with wavelengths longer than 7.96 nm can pass into the coarse-scaled regions with a reflection of less than $5$\%.
        These results are consistent with spectral energy density plots obtained in previous works which use the CAC method for phonon heat transport and the prediction of phonon properties \cite{xiong2014prediction,chen2017ballistic,chen2018passing}.
        
\begin{figure}[h]
            \centering
            \begin{subfigure}{0.48\textwidth}
                \includegraphics[width=\textwidth]{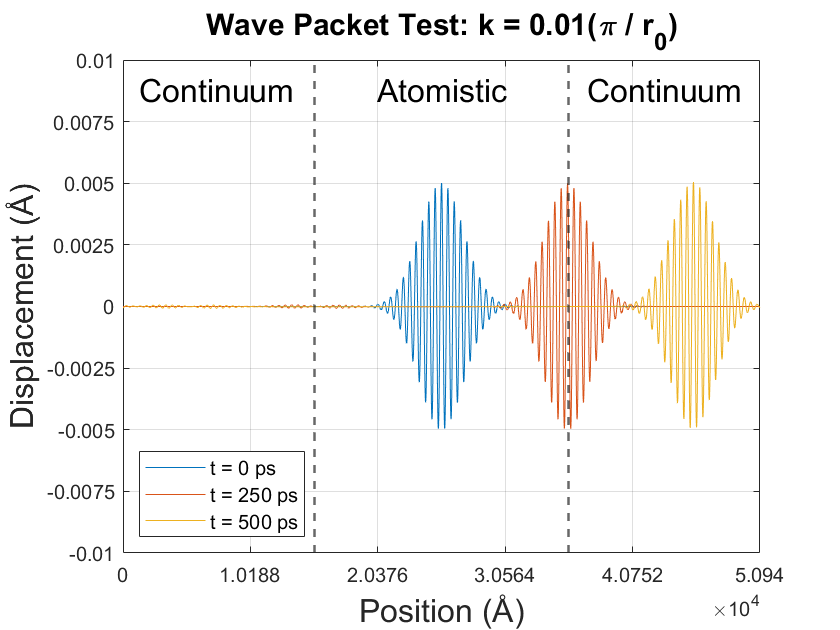}
                \caption{}
            \end{subfigure}
            \begin{subfigure}{0.48\textwidth}
                \includegraphics[width=\textwidth]{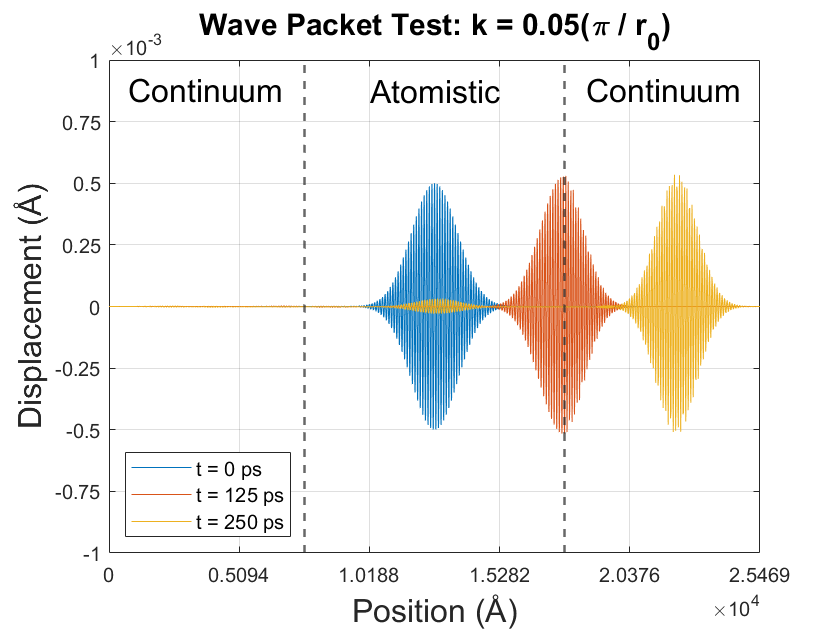}
                \caption{}
            \end{subfigure}
            \\
            \begin{subfigure}{0.48\textwidth}
                \includegraphics[width=\textwidth]{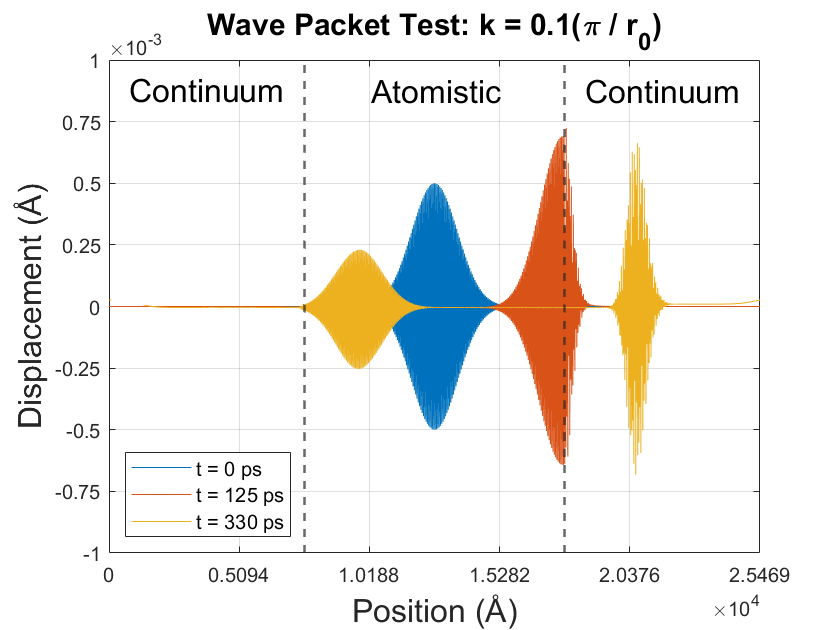}
                \caption{}
            \end{subfigure}
            \begin{subfigure}{0.48\textwidth}
                \includegraphics[width=\textwidth]{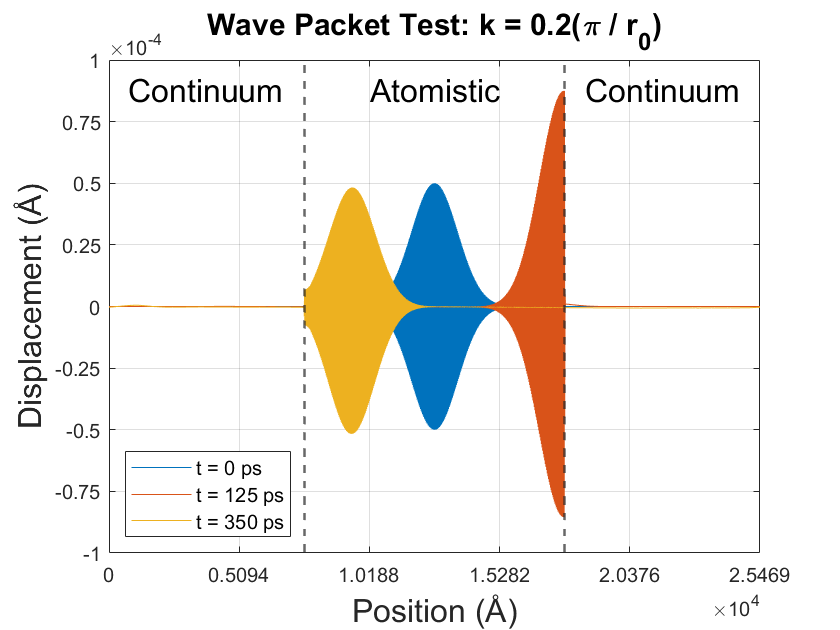}
                \caption{}
            \end{subfigure}
            \caption{Phonon wave packet simulations performed with the following wavevectors: (a) $0.01 \, \pi/r_0$, (b) $0.05 \, \pi/r_0$, (c) $0.1 \, \pi/r_0$, and (d) $0.2 \, \pi/r_0$ \cite{davis2022moving}.}
            \label{Fig:WPTMorse}
        \end{figure}

        Phonon wave packet simulations from previous studies have confirmed that the reflections at the A-C interface are a direct result of the numerical discrepancy between the atomistic and continuum regions \cite{xiong2014prediction,chen2018passing,davis2022moving}.
        An example of this reflection phenomena can be seen in Fig. \ref{Fig:WPTMorse}.
        This mismatch is attributed to the dispersive nature of the frequency-wavevector relation which comes from the fact that the nonlocal internal force-displacement relationship is the only constitutive relation in CAC \cite{xiong2011coarse}.
        Hence, the coarse-scaled regions in CAC simulations impede elastic waves with wavelengths shorter than $2\pi/k_C$.
        To allow these high-frequency waves to pass smoothly from the atomistic to the continuum region, the CAC finite element formulation needs to be modified to allow the full population of phonon waves to propagate across the A-C interface.
    
    \section{Lattice dynamics finite element formulation} \label{Sec: Lattice dynamics-based finite element formulation}
        
        \subsection{Lattice dynamics method} \label{Sec: Lattice Dynamics formulation}
            In this section, we present a technique that was first formulated in \cite{chen2018passing} to overcome the issue of spurious wave reflections at the A-C interfaces, and we add extra details where necessary.
            If we consider a typical polyatomic crystalline system with $N^{\alpha}$ particles in each unit cell, then the standard approximation of the displacement field is given by Eq. (\ref{Eq: Approximate Displacement Field}).
            However, the LD-based method modifies this equation such that the particle displacements are now approximated as follows:
            \begin{equation} \label{Eq: LD_Displacement}
                \textbf{u}^{\alpha}(\textbf{x}, t) = \sum_{j=1}^{2^d} \boldsymbol{\Phi}_j(\textbf{x}) \left[\textbf{U}_j^{\alpha}(t) - \textbf{U}_{sj}^{\alpha}(t) \right] + \textbf{u}_s^{\alpha}(\textbf{x}, t) \, \, \, \, \, \, \, \, \, \, \alpha = 1, 2, \ldots, N^{\alpha}.
            \end{equation}
            In this equation, $\textbf{u}^{\alpha}(\textbf{x}, t)$ is the new displacement at time \textit{t} of the $\alpha^{th}$ atom within a given unit cell located at position $\textbf{x}$; $d$ is the dimensionality of the system; $2^d$ is the total number of nodes in an element; $\boldsymbol{\Phi}_j(\textbf{x})$ is the conventional tri-linear shape function; $\textbf{U}_j^{\alpha}(t)$ is the total displacement of the $\alpha^{th}$ atom in the $j^{th}$ element node at time $t$; $\textbf{U}_{sj}^{\alpha}(t)$ is the short-wavelength displacement (denoted by the subscript $s$) of the $\alpha^{th}$ atom embedded in the $j^{th}$ element node at time $t$; and $\textbf{u}_s^{\alpha}(\textbf{x}, t)$ is the short-wavelength displacement at time $t$ of the $\alpha^{th}$ atom within a unit cell at any material point $\textbf{x}$ (not necessarily a nodal position). 
            Since the tri-linear shape functions satisfy partition of unity $\left(\sum_{j=1}^{2^d}\boldsymbol{\Phi}_j(\textbf{x}) = 1 \right)$, we can rewrite Eq. (\ref{Eq: LD_Displacement}) as follows:
            \begin{equation} \label{Eq: LD_Displacement_2}
                \textbf{u}^{\alpha}(\textbf{x}, t) = \sum_{j=1}^{2^d} \boldsymbol{\Phi}_j(\textbf{x}) \left[\textbf{U}_j^{\alpha}(t) - \textbf{U}_{sj}^{\alpha}(t) + \textbf{u}_s^{\alpha}(\textbf{x}, t) \right].
            \end{equation}
            As a result of this new basis, the CAC governing equation must be updated to account for the modified displacement interpolation which is now a function of time:
            \begin{equation} \label{Eq: FEM Short Wavelength Governing Equation} 
                \rho^{\alpha} \Ddot{\textbf{u}}^{\alpha}(\textbf{x}, t) = \textbf{f}_{int}^{\alpha}(\textbf{x}) + \textbf{f}^{\alpha}(\textbf{x}).
            \end{equation}
            
            In Eq. (\ref{Eq: LD_Displacement_2}), the additional components of the displacement field approximation are $\textbf{U}_{sj}^{\alpha}(t)$ and $\textbf{u}_s^{\alpha}(\textbf{x}, t)$ which represent the short-wavelength displacements that need to be calculated.
            We note that at a given node $j$, $\textbf{U}_{sj}^{\alpha}(t) = \textbf{u}_s^{\alpha}(\textbf{x}, t)$.
            Therefore, the displacements of particles at nodal locations remain unchanged when introducing the short-wavelength basis function.
            Instead, only the neighboring particles at non-nodal unit cells located at material points $\textbf{x}$ within an element get modified by Eq. (\ref{Eq: LD_Displacement_2}).
            These enhanced displacements will influence the force calculations at nodal locations which will allow short-wavelength phonons to pass through the coarse-scaled region.
            
            For a harmonic approximation, atomic displacements can be decomposed into a linear combination of normal modes with a discrete set of wavevectors where the number of wavevectors equals the number of unit cells \cite{born1955dynamical,kumar1988lattice}.
            If we only consider the contributions from short-wavelength phonons with $\textbf{k} > \textbf{k}_C$, then the displacement of the $\alpha^{th}$ particle in a unit cell at undeformed position $\textbf{x}$ is given as follows:
            \begin{equation} \label{Eq: Normal Modes Displacement}
                \textbf{u}_s^{\alpha}(\textbf{x}, t) = \frac{1}{(N_l m^{\alpha})^{1/2}} \sum_{\textbf{k}, \nu (\textbf{k} > \textbf{k}_C)} \textbf{e}_{\textbf{k} \nu}^{\alpha} Q_{\textbf{k} \nu} \mathrm{exp} \left[i \left(\textbf{k} \cdot \textbf{x} - \omega_{\textbf{k} \nu} t \right) \right].
            \end{equation}
            In Eq. (\ref{Eq: Normal Modes Displacement}), each linear combination of normal modes represents the contribution from a wave with wavevector $\textbf{k}$ and phonon branch $\nu$.
            Additionally, $N_l$ is the total number of unit cells in the system; $m^{\alpha}$ is the mass of the $\alpha^{th}$ particle in the $l^{th}$ unit cell; $\textbf{e}_{\textbf{k} \nu}^{\alpha}$ is the polarization vector that determines which direction each particle moves; $Q_{\textbf{k} \nu}$ is the normal mode coordinate which gives both the amplitude of the wave and the time dependence; and $\omega_{\textbf{k} \nu}$ is the angular frequency corresponding to wavevector $\textbf{k}$.
            We can then rewrite Eq. (\ref{Eq: Normal Modes Displacement}) to obtain the following expressions for $\textbf{U}_{sj}^{\alpha}(t)$ and $\textbf{u}_s^{\alpha}(\textbf{x}, t)$ \cite{chen2018passing}:
            \begin{align}
                \textbf{U}_{sj}^{\alpha}(t) &= \frac{1}{N_A} \sum_{\textbf{k}, \nu (\textbf{k} > \textbf{k}_C)} \textbf{e}_{\textbf{k} \nu}^{\alpha} U_{\textbf{k} \nu}^{\alpha} \mathrm{exp} \left[i \left(\textbf{k} \cdot \textbf{x}_j - \omega_{\textbf{k} \nu} t \right) \right] \label{Eq: UsiAlpha} \\
                \textbf{u}_{s}^{\alpha}(\textbf{x}, t) &= \frac{1}{N_A} \sum_{\textbf{k}, \nu (\textbf{k} > \textbf{k}_C)} \textbf{e}_{\textbf{k} \nu}^{\alpha} U_{\textbf{k} \nu}^{\alpha} \mathrm{exp} \left[i \left(\textbf{k} \cdot \textbf{x} - \omega_{\textbf{k} \nu} t \right) \right] \label{Eq: usAlpha}
            \end{align}
            where $N_A$ represents the total number of unit cells in only the \textit{atomistic} region, and $U_{\textbf{k} \nu}^{\alpha}$ is the amplitude.
            The short-wavelength displacement at an unknown position $\textbf{x}$ and time \textit{t} in the coarse-scaled region is linked to information at a known position $\textbf{x}_0$ and time $t_0$ in the fine-scaled region as follows:
            \begin{align}
                \textbf{u}_{s}^{\alpha}(\textbf{x}, t) &= \frac{1}{N_A} \sum_{\textbf{k}, \nu (\textbf{k} > \textbf{k}_C)} \textbf{e}_{\textbf{k} \nu}^{\alpha} U_{\textbf{k} \nu}^{\alpha}(\textbf{x}_0, t_0) \mathrm{exp} \left \{i \left[\textbf{k} \cdot (\textbf{x} - \textbf{x}_0) - \omega_{\textbf{k} \nu} (t - t_0) \right] \right \} \nonumber \\
                &= \frac{1}{N_A} \sum_{\textbf{k}, \nu (\textbf{k} > \textbf{k}_C)} \textbf{e}_{\textbf{k} \nu}^{\alpha} U_{\textbf{k} \nu}^{\alpha}(\textbf{x}_0, t_0) \mathrm{exp} \left [i \left(\textbf{k} \cdot \Delta \textbf{x} - \omega_{\textbf{k} \nu} \Delta t \right) \right ] \label{Eq: usAlpha_2}.
            \end{align}
            Here, we have only shown the expression for $\textbf{u}_{s}^{\alpha}(\textbf{x}, t)$ as the expression for $\textbf{U}_{sj}^{\alpha}(t)$ would have the same form.
            In Eq. (\ref{Eq: usAlpha_2}), $\Delta \textbf{x} = \textbf{x} - \textbf{x}_0$ represents the spatial distance between the current unit cell at location \textbf{x} in the continuum region and the reference unit cell at \textit{undeformed} location $\textbf{x}_0$ in the atomistic region.
            Additionally, $\Delta t = t - t_0$ represents the difference between the current time \textit{t} and the time $t_0$ at which $U_{\textbf{k} \nu}^{\alpha}(\textbf{x}_0, t_0)$ was calculated.
            
            We can use Eq. (\ref{Eq: usAlpha_2}) to calculate $\textbf{u}_{s}^{\alpha}(\textbf{x}, t)$ [and $\textbf{U}_{sj}^{\alpha}(t)$] and then substitute these expressions into Eq. (\ref{Eq: LD_Displacement_2}).
            As a result, short-wavelength effects will now be incorporated into $\textbf{u}^{\beta}(\textbf{x}')$ from Eq. (\ref{Eq: Internal Force Density}) which will update the internal force calculation.
            Specifically, the forces at the nodes will now contain information from the entire spectrum of phonon waves: low-frequency data from linear interpolation and high-frequency data from LD calculations.
            Therefore, it is clear that an accurate determination of $\textbf{u}_{s}^{\alpha}(\textbf{x}, t)$ and $\textbf{U}_{sj}^{\alpha}(t)$ is crucial to achieve proper force matching, and this requires calculating the amplitude $U_{\textbf{k} \nu}^{\alpha}(\textbf{x}_0, t_0)$ of each short-wavelength phonon mode.
            We derive this amplitude in the following section.
        
        \subsection{Determining the amplitude of the short-wavelength phonon mode} \label{Sec: Determining the amplitude of the short-wavelength phonon mode}
            We can represent the short-wavelength displacement of the $\alpha^{th}$ particle at undeformed position $\textbf{x}_j$ and time $t$ as follows:
            \begin{equation} \label{Eq: Short wavelength amplitude}
                \textbf{u}_{s}^{\alpha}(\textbf{x}_j, t) = \sum_{\textbf{k}, \nu} A_{\textbf{k} \nu}^{\alpha} \textbf{e}_{\textbf{k} \nu}^{\alpha} \mathrm{exp} \left [i \left(\textbf{k} \cdot \textbf{x}_j - \omega_{\textbf{k} \nu} t \right) \right] + B_{\textbf{k} \nu}^{\alpha} \textbf{e}_{\textbf{k} \nu}^{\alpha} \mathrm{exp} \left [i \left(\textbf{k} \cdot \textbf{x}_j + \omega_{\textbf{k} \nu} t \right) \right].
            \end{equation}
           As before, we have only shown the expression for $\textbf{u}_{s}^{\alpha}(\textbf{x}_j, t)$ as the same analysis applies to $\textbf{U}_{sj}^{\alpha}(t)$.
           Eq. (\ref{Eq: Short wavelength amplitude}) is a general expression for the short-wavelength displacement, but it is understood that $\textbf{x}_j = \textbf{x}_0$ and $t = t_0$ in this example.
           Here, $A_{\textbf{k} \nu}^{\alpha}$ and $B_{\textbf{k} \nu}^{\alpha}$ are the two unknown coefficients computed for each mode which represent both parts of the coefficient $U_{\textbf{k} \nu}^{\alpha}(\textbf{x}_0, t_0)$.
           Hence, the goal is to calculate $A_{\textbf{k} \nu}^{\alpha}$ and $B_{\textbf{k} \nu}^{\alpha}$ at $t_0$ as these coefficients will then be applied to the short-wavelength calculation (Eq. \ref{Eq: usAlpha_2}) at every subsequent time step.
           
           To find these amplitudes, we must take the discrete Fourier transform (DFT) of both the initial displacements $\textbf{u}_{j}^{\alpha}$ and initial velocities $\textbf{v}_{j}^{\alpha}$ in the atomistic region as shown below:
           \begin{align}
               \textbf{C}_{\textbf{k}}^{\alpha} &= \sum_{j=0}^{N_A-1} \textbf{u}_j^{\alpha} \mathrm{exp} \left[-i \left(\textbf{k} \cdot \textbf{x}_j \right) \right] \label{Eq: DFT Displacement} \\
               \textbf{D}_{\textbf{k}}^{\alpha} &= \sum_{j=0}^{N_A-1} \textbf{v}_j^{\alpha} \mathrm{exp} \left[-i \left(\textbf{k} \cdot \textbf{x}_j \right) \right]. \label{Eq: DFT Velocity}
           \end{align}
           where $\textbf{x}_j = j\textbf{r}_0$ is the position of the $j^{th}$ unit cell in the undeformed configuration with $\mathbf{r}_0$ being the equilibrium spacing.
           We can then relate the modal amplitude in Eq. (\ref{Eq: DFT Displacement}) to the phonon modes in Eq. (\ref{Eq: Short wavelength amplitude}) evaluated at $t = 0$ for a specific wavevector \textbf{k}:
           \begin{align}
               \textbf{C}_{\textbf{k}}^{\alpha} &= \sum_{\nu} \textbf{u}_{s}^{\alpha}(\textbf{x}_j, 0) \mathrm{exp} \left[-i (\textbf{k} \cdot \textbf{x}_j) \right] \nonumber \\
               &= \sum_{\nu} A_{\textbf{k} \nu}^{\alpha} \textbf{e}_{\textbf{k} \nu}^{\alpha} \mathrm{exp} \left [i \left(\textbf{k} \cdot \textbf{x}_j - \textbf{k} \cdot \textbf{x}_j - \omega_{\textbf{k} \nu} 0 \right) \right] + B_{\textbf{k} \nu}^{\alpha} \textbf{e}_{\textbf{k} \nu}^{\alpha} \mathrm{exp} \left [i \left(\textbf{k} \cdot \textbf{x}_j - \textbf{k} \cdot \textbf{x}_j + \omega_{\textbf{k} \nu} 0 \right) \right] \nonumber \\
               &= \sum_{\nu} \left(A_{\textbf{k} \nu}^{\alpha} + B_{\textbf{k} \nu}^{\alpha} \right) \textbf{e}_{\textbf{k} \nu}^{\alpha}.
           \end{align}
           Next, we can perform a similar analysis for the modal amplitude of the velocities in Eq. (\ref{Eq: DFT Velocity}) by taking the derivative of Eq. (\ref{Eq: Short wavelength amplitude}) with respect to $t$:
           \begin{align}
               \textbf{D}_{\textbf{k}}^{\alpha} &= \sum_{\nu} \textbf{v}_{s}^{\alpha}(\textbf{x}_j, 0) \mathrm{exp} \left[-i (\textbf{k} \cdot \textbf{x}_j) \right] \nonumber \\
               &= \sum_{\nu} -i \omega_{\textbf{k} \nu} A_{\textbf{k} \nu}^{\alpha} \textbf{e}_{\textbf{k} \nu}^{\alpha} \mathrm{exp} \left [i \left(\textbf{k} \cdot \textbf{x}_j - \textbf{k} \cdot \textbf{x}_j - \omega_{\textbf{k} \nu} 0 \right) \right] + i \omega_{\textbf{k} \nu} B_{\textbf{k} \nu}^{\alpha} \textbf{e}_{\textbf{k} \nu}^{\alpha} \mathrm{exp} \left [i \left(\textbf{k} \cdot \textbf{x}_j - \textbf{k} \cdot \textbf{x}_j + \omega_{\textbf{k} \nu} 0 \right) \right] \nonumber \\
               &= \sum_{\nu} \left(B_{\textbf{k} \nu}^{\alpha} - A_{\textbf{k} \nu}^{\alpha} \right) i \omega_{\textbf{k} \nu} \textbf{e}_{\textbf{k} \nu}^{\alpha}.
           \end{align}
           Finally, we arrive at a system of two equations with the two unknowns $A_{\textbf{k} \nu}^{\alpha}$ and $B_{\textbf{k} \nu}^{\alpha}$:
           \begin{align}
               \textbf{C}_{\textbf{k}}^{\alpha} &= \sum_{j=0}^{N_A-1} \textbf{u}_j^{\alpha} \mathrm{exp} \left[-i \left(\textbf{k} \cdot \textbf{x}_j \right) \right] = \sum_{\nu} \left(A_{\textbf{k} \nu}^{\alpha} + B_{\textbf{k} \nu}^{\alpha} \right) \textbf{e}_{\textbf{k} \nu}^{\alpha} \label{Eq: Modal amplitude displacements} \\
               \textbf{D}_{\textbf{k}}^{\alpha} &= \sum_{j=0}^{N_A-1} \textbf{v}_j^{\alpha} \mathrm{exp} \left[-i \left(\textbf{k} \cdot \textbf{x}_j \right) \right] = \sum_{\nu} \left(B_{\textbf{k} \nu}^{\alpha} - A_{\textbf{k} \nu}^{\alpha} \right) i \omega_{\textbf{k} \nu} \textbf{e}_{\textbf{k} \nu}^{\alpha}. \label{Eq: Modal amplitude velocities}
           \end{align}
           Therefore, the DFTs of displacement and velocity for the $\alpha^{th}$ particle within each unit cell produce a $2\nu$ by $2\nu$ matrix to solve for the coefficients $A_{\textbf{k} \nu}^{\alpha}$ and $B_{\textbf{k} \nu}^{\alpha}$ corresponding to a given wavevector $\textbf{k}$.
           We solve these equations for a one-dimensional monatomic chain in \ref{Sec: Solving for the short-wavelength amplitude in 1D}. 
    
        \subsection{Passing a single high-frequency wave packet from atomistic to continuum} \label{Sec: Passing single high-frequency waves from atomistic to continuum}
\begin{figure}[htpb]
                \centering
                \begin{subfigure}{0.48\textwidth}
                    \includegraphics[width=\textwidth]{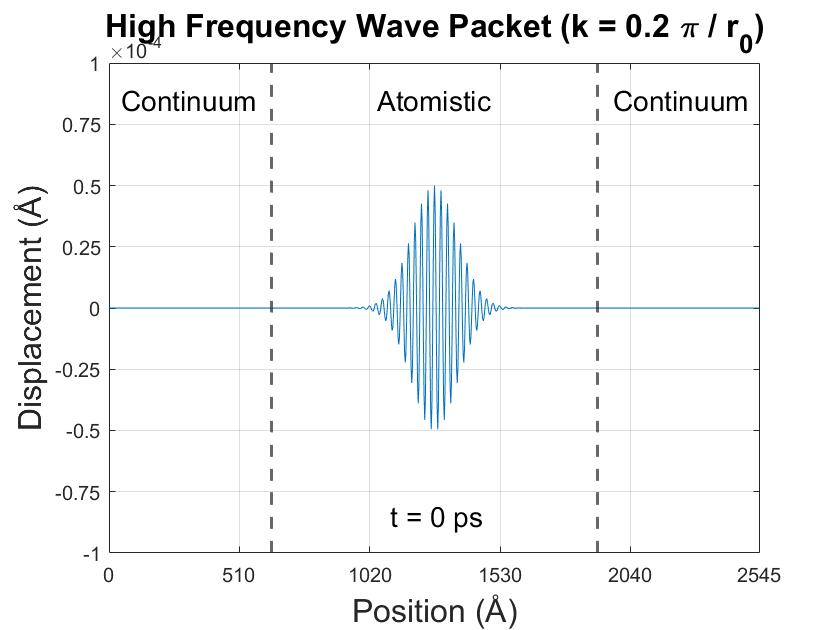}
                    \caption{}
                \end{subfigure}
                \begin{subfigure}{0.48\textwidth}
                    \includegraphics[width=\textwidth]{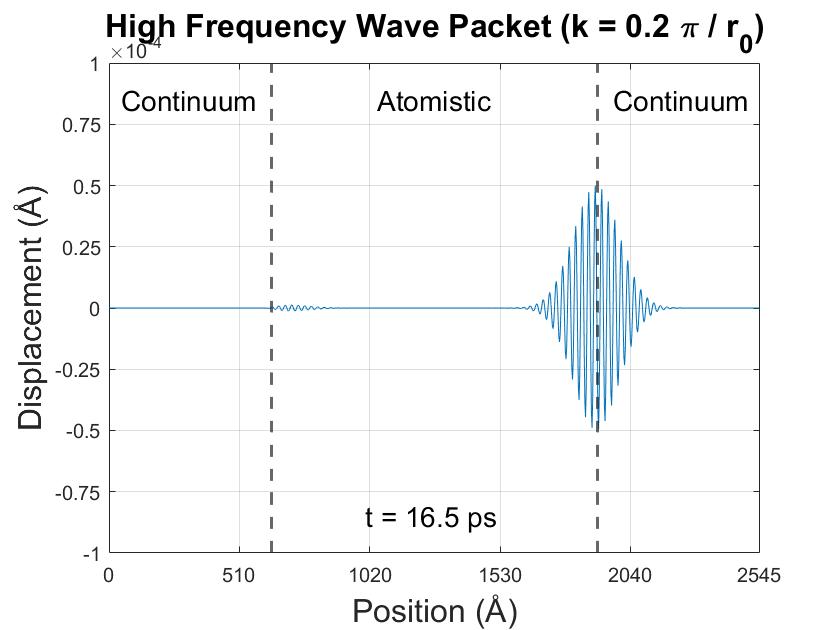}
                    \caption{}
                \end{subfigure}
                \\
                \begin{subfigure}{0.48\textwidth}
                    \includegraphics[width=\textwidth]{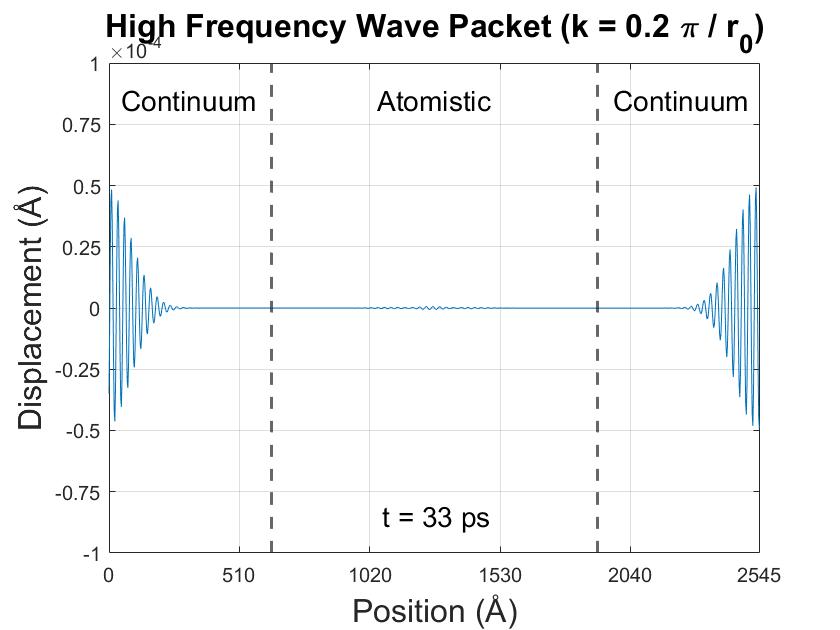}
                    \caption{}
                \end{subfigure}
                \begin{subfigure}{0.48\textwidth}
                    \includegraphics[width=\textwidth]{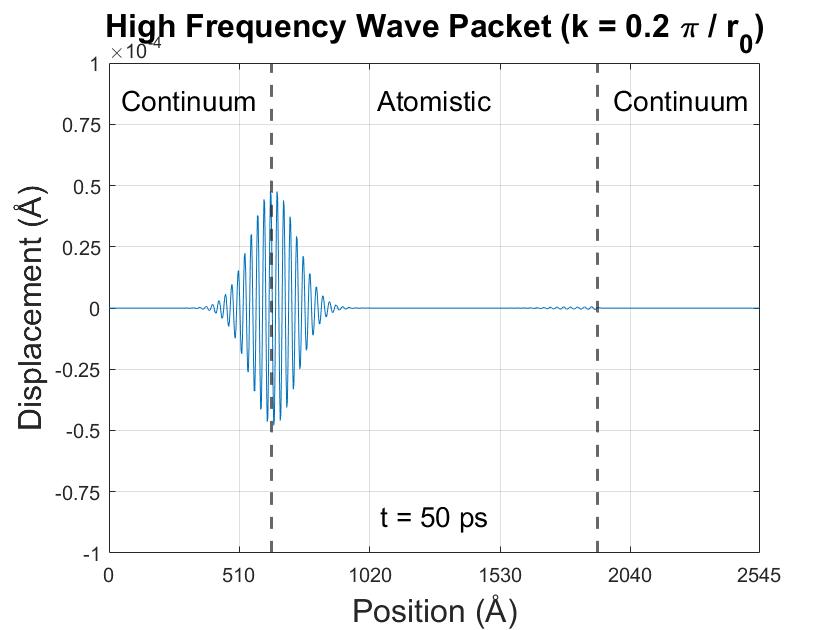}
                    \caption{}
                \end{subfigure}
                \caption{High-frequency phonon wave packet simulation performed with the LD interpolation method ($k = 0.2 \, \pi/r_0$).}
                \label{Fig:LDWavePacketSimulation}
            \end{figure}

            We now present a 1D wave packet simulation performed with the LD interpolation method.
            The results can be seen in Fig. \ref{Fig:LDWavePacketSimulation}, and $k = 0.2 \, \pi/r_0$ in this case.
            Hence, we can directly compare the results in Fig. \ref{Fig:LDWavePacketSimulation} to the results in Fig. \ref{Fig:WPTMorse}d.
            We observe that the LD interpolation scheme permits the entire phonon wave packet to travel across the A-C interface from the atomistic to the continuum region with a $>$99.5\% transmission.
            This is in contrast to the complete reflection seen in Fig. \ref{Fig:WPTMorse}d and is congruent with the results from previous studies \cite{chen2018passing}.
            Additionally, by enabling periodic boundary conditions, we observe that the LD interpolation method allows the high-frequency phonon wave packet to travel between the two outer continuum regions and back to the center atomistic region.
            The transmission demonstrated in Fig. \ref{Fig:LDWavePacketSimulation} validates the implementation of the LD interpolation method.
    
    \section{Lattice dynamics technique for multiple waves} \label{Sec: Upgrading the lattice dynamics technique for multiple waves}
    
        \subsection{Background and preliminary approach} \label{Sec: Background and preliminary approach}
            While the method presented in the previous section has been shown to efficiently pass high-frequency phonons between the atomistic and continuum regions of a CAC domain, the scheme is limited to single wave packets of a specified wavevector. 
            This is because the short-wavelength amplitude information can only be stored for one wave at a time to prevent data from being overwritten.
            In this section, we present an LD interpolation method to be used with multiple waves and wavevectors in a single CAC domain. 
            We recall the expression for the displacement field approximation from Sec. \ref{Sec: Lattice Dynamics formulation}:
            \begin{equation} \label{Eq: LD_Displacement_3}
                \textbf{u}^{\alpha}(\textbf{x}, t) = \sum_{j=1}^{2^d} \boldsymbol{\Phi}_j(\textbf{x}) \left[\textbf{U}_j^{\alpha}(t) - \textbf{U}_{sj}^{\alpha}(t) + \textbf{u}_s^{\alpha}(\textbf{x}, t) \right].
            \end{equation}
            Here, we note that both $\textbf{U}_{sj}^{\alpha}(t)$ and $\textbf{u}_s^{\alpha}(\textbf{x}, t)$ contain all the short-wavelength information of a given wave packet at time $t$.
            Again, since the same analysis applies to both terms, we only focus on $\textbf{u}_s^{\alpha}(\textbf{x}, t)$ in this section.
            
\begin{figure}[htpb]
                \centering
                \begin{subfigure}{0.48\textwidth}
                    \includegraphics[width=\textwidth]{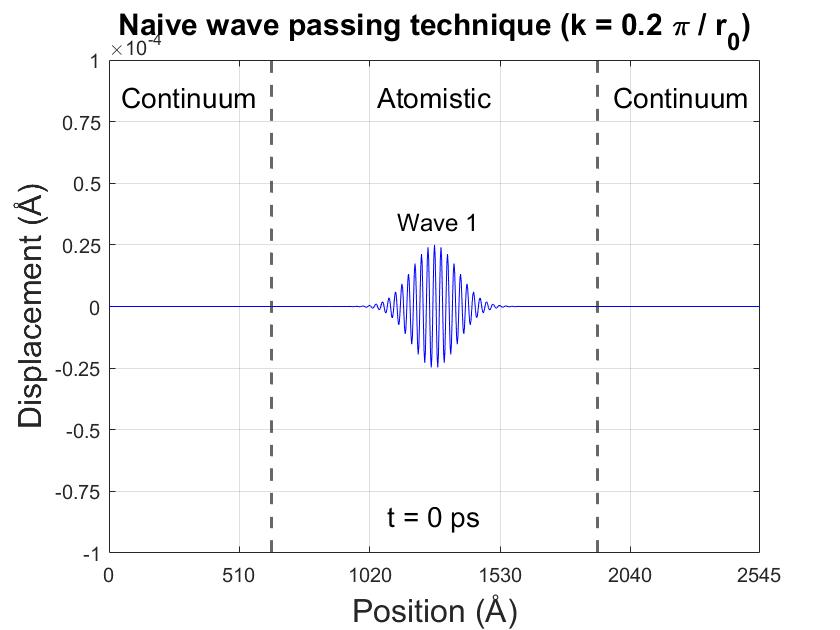}
                    \caption{}
                \end{subfigure}
                \begin{subfigure}{0.48\textwidth}
                    \includegraphics[width=\textwidth]{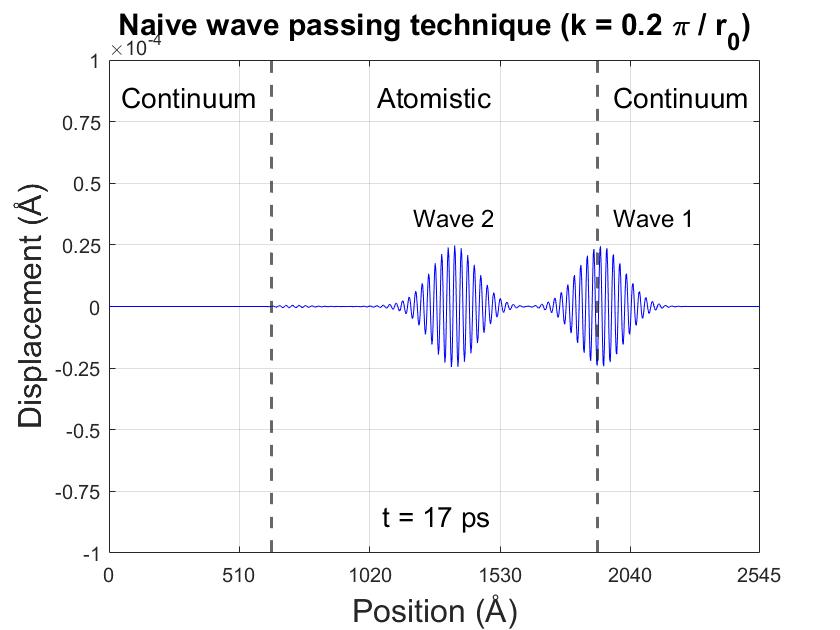}
                    \caption{}
                \end{subfigure}
                \\
                \begin{subfigure}{0.48\textwidth}
                    \includegraphics[width=\textwidth]{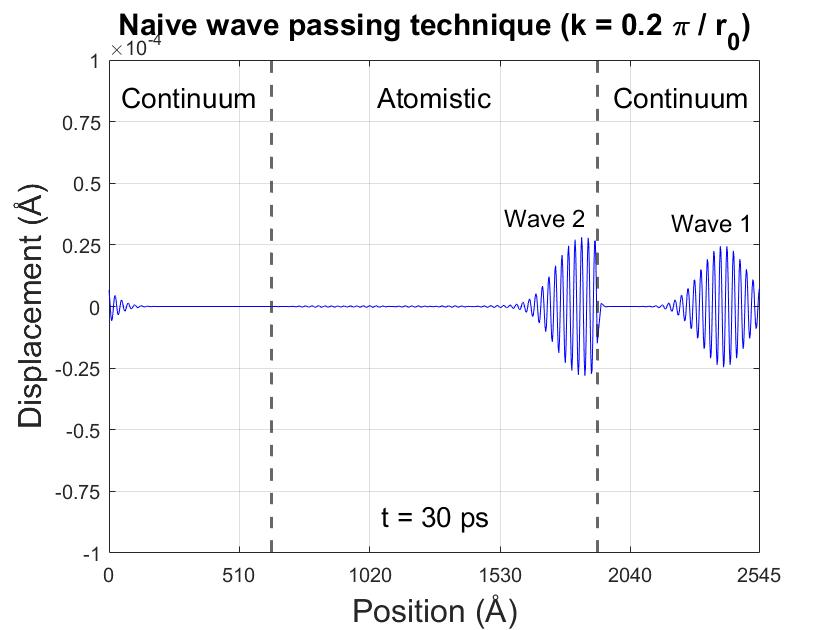}
                    \caption{}
                \end{subfigure}
                \caption{Naive approach to passing multiple waves across the A-C interface in which wave 1 crosses but wave 2 gets reflected ($k = 0.2 \, \pi/r_0$).}
                \label{Fig:NaiveWavePassingTechnique}
            \end{figure}

            For multiple wave packets, a straight-forward extension to Eq. (\ref{Eq: Short wavelength amplitude}) is given by
            \begin{align} \label{Eq: Short wavelength amplitude multiple waves}
                \textbf{u}_{s}^{\alpha}(\textbf{x}_j, t) &= \sum_{\textbf{k}, \nu} \sum_l A_{\textbf{k} \nu, l}^{\alpha} \textbf{e}_{\textbf{k} \nu}^{\alpha} \mathrm{exp} \left \{i \left[\textbf{k} \cdot \textbf{x}_j - \omega_{\textbf{k} \nu} (t - t_l) \right] \right\} \\ \nonumber
                &+ B_{\textbf{k} \nu, l}^{\alpha} \textbf{e}_{\textbf{k} \nu}^{\alpha} \mathrm{exp} \left \{i \left[\textbf{k} \cdot \textbf{x}_j + \omega_{\textbf{k} \nu} (t - t_l) \right] \right\}
            \end{align}
            where the inner summation occurs over all wave packets $l$ nucleated at time $t_l$.
            The coefficients $A_{\textbf{k} \nu, l}^{\alpha}$ and $B_{\textbf{k} \nu, l}^{\alpha}$ described in Sec. \ref{Sec: Determining the amplitude of the short-wavelength phonon mode} now correspond to each wave packet $l$. 
            The problem with this approach is that except for the first wave packet, the other wave packets generated at time $t_l$ cannot be tracked over time.
            As a result, any new phonon initialized at time $t_l \ne 0$ will be reflected off the A-C interface as seen in Fig. \ref{Fig:NaiveWavePassingTechnique}.
            Specifically, the main issue lies in keeping track of each individual phonon generated at time $t_l$ without losing information from other phonons.
            In an attempt to overcome this difficult problem, we provide a detailed solution below.
            
        \subsection{Solution to the preliminary approach} \label{Sec: Solution to the preliminary approach}
            Each wave packet can be characterized by its wavevector and frequency combination ($\textbf{k}$, $\omega$), its short-wavelength amplitudes ($A_{\textbf{k} \nu,l}^{\alpha}$ and $B_{\textbf{k} \nu,l}^{\alpha}$), and the time at which it was initialized ($t_l$).
            These terms must be tracked and stored correctly in order to allow multiple waves to pass across the A-C interfaces. 
            To this end, we first rewrite Eq. (\ref{Eq: Short wavelength amplitude multiple waves}) as follows:
            \begin{align} \label{Eq: Short wavelength amplitude multiple waves 2}
                \textbf{u}_{s}^{\alpha}(\textbf{x}_j, t) &= \sum_{\textbf{k}, \nu} \sum_l A_{\textbf{k} \nu, l}^{\alpha} \mathrm{exp}[i \cdot \omega_{\textbf{k} \nu} t_l] \textbf{e}_{\textbf{k} \nu}^{\alpha} \mathrm{exp} \left [i \left(\textbf{k} \cdot \textbf{x}_j - \omega_{\textbf{k} \nu} t \right) \right] \\ \nonumber
                &+ B_{\textbf{k} \nu, l}^{\alpha} \mathrm{exp}[- i \cdot \omega_{\textbf{k} \nu} t_l] \textbf{e}_{\textbf{k} \nu}^{\alpha} \mathrm{exp} \left [i \left(\textbf{k} \cdot \textbf{x}_j + \omega_{\textbf{k} \nu} t \right) \right].
            \end{align}
            Thus, we now have ``time-stamped" coefficients $A_{\textbf{k} \nu, l}^{\alpha} \mathrm{exp}[i \cdot \omega_{\textbf{k} \nu} t_l]$ and $B_{\textbf{k} \nu, l}^{\alpha} \mathrm{exp}[- i \cdot \omega_{\textbf{k} \nu}  t_l]$ which contain the unique information for each phonon and encode the time at which the wave is nucleated.
            During each time step (before a new wave is generated), we take a ``snapshot" of the domain in \textit{\textbf{k}-space} whereby the amplitude coefficients are obtained from the Fourier transform discussed in Sec. \ref{Sec: Determining the amplitude of the short-wavelength phonon mode}.
            After the generation of a new phonon, a second Fourier transform of the domain is taken, and the first set of coefficients is subtracted from the second.
            This allows us to see which frequencies are ``new" and thus gives us information about the current wave without the influence from previous phonons.
            Finally, we add the difference in these coefficients to a global ``master" array and use a modified form of Eq. (\ref{Eq: usAlpha_2}) to calculate the displacement of each particle:
            \begin{align}
                \textbf{u}_{s}^{\alpha}(\textbf{x}, t) = \frac{1}{N_A} \sum_{\textbf{k}, \nu (\textbf{k} > \textbf{k}_C)} \sum_l \textbf{e}_{\textbf{k} \nu}^{\alpha} U_{\textbf{k} \nu, l}^{\alpha} \mathrm{exp}[i \cdot \omega_{\textbf{k} \nu} t_l] \mathrm{exp} \left [i \left(\textbf{k} \cdot \Delta \textbf{x} - \omega_{\textbf{k} \nu} \Delta t \right) \right ] \label{Eq: usAlpha_3}.
            \end{align}
            As a result, the displacement field approximation is updated based upon multiple waves, and no information gets lost.
            
        \subsection{Detailed explanation in 1D} \label{Sec: Detailed explanation in 1D}
\begin{figure}[htpb]
                \centering
                \includegraphics[width=0.75\textwidth]{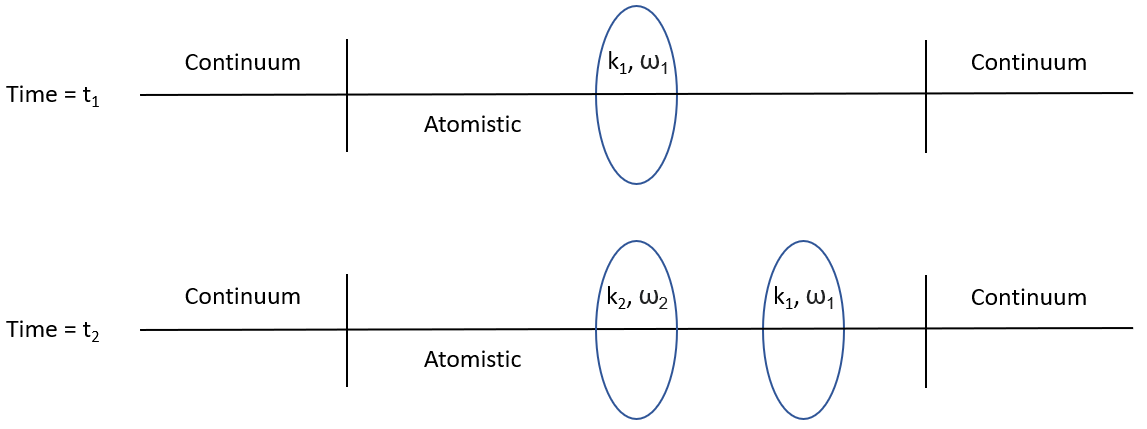}
                \caption{Schematic of two phonon wave packets traveling through a CAC domain.}
                \label{Fig:WavePacketSchematic}
            \end{figure}

            We now elaborate on this process for a one-dimensional monatomic chain as is utilized in the present work.
            Fig. \ref{Fig:WavePacketSchematic} gives a visual representation of two high-frequency wave packets with wavevector-frequency pairs of ($k_1$, $\omega_1$) and ($k_2$, $\omega_2$) traveling within the 1D CAC framework described in Sec. \ref{Sec: Domain geometry}.
            The first phonon is generated at time $t_1$, the second phonon is generated at time $t_2$, and without loss of generality, we assume that each wave originates at the center of the atomistic region.
            For a one-dimensional system, Eq. (\ref{Eq: Short wavelength amplitude multiple waves 2}) reduces to the following:
            \begin{align} \label{Eq: Short wavelength amplitude multiple waves 1D}
                u_{s}(x_j, t) &= \sum_k \sum_l A_{k, l} \mathrm{exp}[i \cdot \omega_{k} t_l] \mathrm{exp} \left [i \left(k \cdot x_j - \omega_{k} t \right) \right] \\ \nonumber
                &+ B_{k, l} \mathrm{exp}[- i \cdot \omega_{k} t_l] \mathrm{exp} \left [i \left(k \cdot x_j + \omega_{k} t \right) \right].
            \end{align}
            By following the procedure discussed in Sec. \ref{Sec: Determining the amplitude of the short-wavelength phonon mode}, we can solve for the coefficients and substitute these back into Eq. (\ref{Eq: usAlpha_2}) to achieve the following short-wavelength displacement approximation in 1D:
            \begin{equation} \label{Eq: Short wavelength displacement multiple waves 1D exp}
                u_{s}(x, t) = \frac{1}{N_A} \sum_{k (k > k_C)} \sum_l C_{k,l} \mathrm{exp}[i \cdot \omega_{k} t_l] \mathrm{exp} \left [i \left(k \cdot x - \omega_{k} t \right) \right]
            \end{equation}
            where $t$ is the global simulation time, and $C_{k,l}$ is the derived coefficient given by Eq. (\ref{Eq: Ck equation exponential}).
            Recall that $C_{k,l}$ is purely a function of the atomic displacements, undeformed positions, and wavevectors.
            Furthermore, Eq. (\ref{Eq: Short wavelength displacement multiple waves 1D exp}) is the same as Eq. (\ref{Eq: us_1D_exp}) but with the added exponential term and summation over $l$.
        
            Therefore, we have the new \textit{time-stamped} coefficient $E_{k,l} = C_{k,l} \mathrm{exp}[i \cdot \omega_{k} t_l]$. 
            Expanding out $E_{k,l}$ into its real and imaginary parts, we get the following:
            \begin{equation}
                E_{k,l} = C_{k,l} \mathrm{exp}[i \cdot \omega_{k} t_l] = \left[\mathrm{Re}(C_{k,l}) - i \mathrm{Im}(C_{k,l}) \right] \left[\mathrm{cos}(\omega_k t_l) + i \mathrm{sin}(\omega_k t_l) \right]
            \end{equation}
            where Re($C_{k,l}$) and Im($C_{k,l}$) are given by Eqs. (\ref{Eq: Ck_real_coefficient}) and (\ref{Eq: Ck_imag_coefficient}) respectively.
            Next, we can define the real and imaginary components of the coefficient $E_{k,l}$:
            \begin{align}
                \mathrm{Re}(E_{k,l}) &= \mathrm{Re}(C_{k,l}) \mathrm{cos}(\omega_k t_l) + \mathrm{Im}(C_{k,l}) \mathrm{sin}(\omega_k t_l) \label{Eq: E_k_real} \\
                \mathrm{Im}(E_{k,l}) &= \mathrm{Re}(C_{k,l}) \mathrm{sin}(\omega_k t_l) - \mathrm{Im}(C_{k,l}) \mathrm{cos}(\omega_k t_l). \label{Eq: E_k_imag}
            \end{align}
            Substituting the two parts of this coefficient back into Eq. (\ref{Eq: Short wavelength displacement multiple waves 1D exp}) and writing the expression in trigonometric form, we get the following:
            \begin{equation} \label{Eq: Short wavelength displacement multiple waves 1D trig}
                u_{s}(x, t) = \frac{1}{N_A} \sum_{k (k > k_C)} \sum_l \left[\mathrm{Re}(E_{k,l}) + i \mathrm{Im}(E_{k,l}) \right] \left[\mathrm{cos} (k \cdot x - \omega_{k} t) + i \mathrm{sin} (k \cdot x - \omega_{k} t) \right].
            \end{equation}
            Then, keeping only the real parts of Eq. (\ref{Eq: Short wavelength displacement multiple waves 1D trig}), we arrive at the final expression for the multi-wave, short-wavelength displacement in 1D:
            \begin{equation} \label{Eq: Short wavelength displacement multiple waves 1D real trig}
                u_{s}(x, t) = \frac{1}{N_A} \sum_{k (k > k_C)} \sum_l \left[\mathrm{Re}(E_{k,l}) \mathrm{cos} (k \cdot x - \omega_{k} t) - \mathrm{Im}(E_{k,l}) \mathrm{sin}(k \cdot x - \omega_{k} t) \right].
            \end{equation}
            Equation (\ref{Eq: Short wavelength displacement multiple waves 1D real trig}) allows us to update the atomic displacements given multiple high-frequency waves in the CAC domain. 
        
        \subsection{Using the LD technique with time integration} \label{Sec: Using the LD technique with time integration}
            We now discuss how the process described above is incorporated into the time integration algorithm, and we use the two waves from Fig. \ref{Fig:WavePacketSchematic} as a reference.
            Additionally, we assume that $t = t_2$ and the first phonon (wave 1) has already been nucleated in the atomistic region.
            The steps are enumerated as follows.
        
            \begin{enumerate}
                \item After the particle velocity update, we calculate the \textit{time-independent} amplitude coefficients $C_{k,l}$.
                Specifically, we find the real and imaginary components of the coefficient $C_{k,l}$ using Eqs. (\ref{Eq: Ck_real_coefficient}) and (\ref{Eq: Ck_imag_coefficient}) and store them in a \textit{k}-based array in which each index is a different wavevector.
                This effectively allows us to take a ``snapshot" of the framework in \textit{k}-space and thus capture the information from any phonon currently within the domain.
                Referring back to Fig. \ref{Fig:WavePacketSchematic}, we calculate and store the $C_{k,1}$ coefficients to preserve the displacements/velocities induced by wave 1.
            
                \item If desired, we then generate the second phonon (wave 2) after obtaining $C_{k,1}$ and update the particle displacements and velocities accordingly.
                In other words, displacements and velocities resulting from wave 2 are added to those values induced by wave 1 such that both phonons are still present in the domain and information from each is preserved.
            
                \item At the end of the time step, we then calculate the time-independent amplitude coefficient of wave 2 ($C_{k,2}$).
                We note that the wavevector of wave 2 can be any value -- it does not have to be the same as wave 1.
            
                \item  Next, we subtract the real and imaginary components of $C_{k,1}$ from the corresponding components of $C_{k,2}$.
                This gives us the exclusive frequencies from wave 2 as seen below:
                \begin{align}
                    \mathrm{Re}(C_{k,21}) &= \mathrm{Re}(C_{k,2}) - \mathrm{Re}(C_{k,1}) \\
                    \mathrm{Im}(C_{k,21}) &= \mathrm{Im}(C_{k,2}) - \mathrm{Im}(C_{k,1}).
                \end{align}
            
                \item Finally, we substitute $\mathrm{Re}(C_{k,21})$ and $\mathrm{Im}(C_{k,21})$ into Eqs. (\ref{Eq: E_k_real}) and (\ref{Eq: E_k_imag}) to obtain the new \textit{time-stamped} coefficient $E_{k,2}$.
                This coefficient contains all the ``new" information from wave 2 including its generation time.
            \end{enumerate}
        
\begin{figure}[htpb]
                \centering
                \includegraphics[width=0.6\textwidth]{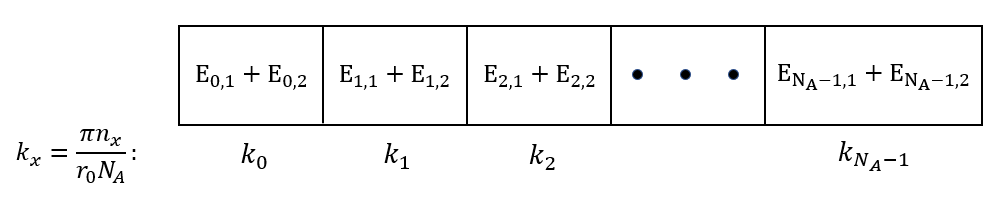}
                \caption{Master template for the time-stamped wave passing coefficients.}
                \label{Fig:MasterTemplate}
            \end{figure}

            The real and imaginary parts of $E_{k,2}$ are added to a global \textit{k}-based array where each array index contains the sum of the $E_{k,l}$ coefficients from every generated phonon (the $E_{k,1}$ coefficients from wave 1 would have already been obtained at $t = t_1$).
            This array serves as a ``master template" by storing the time-stamped coefficients from every phonon, and a visual representation for wave 1 and wave 2 can be seen in Fig. \ref{Fig:MasterTemplate}.
            We note that the nondegenerate wavevectors are limited to $k_x = \pi n_x / r_0 N_A$ where $n_x$ is an integer ranging from $0$ to $N_A-1$ \cite{thomas2010predicting}.
            Thus, for any given wavevector, we know the corresponding total amplitude coefficient.
            We can then use these coefficients in Eq. (\ref{Eq: Short wavelength displacement multiple waves 1D real trig}) during all subsequent time steps to calculate the short wavelength displacement induced by multiple wave packets.
        
\begin{figure}[htpb]
                \centering
                \includegraphics[width=0.95\textwidth]{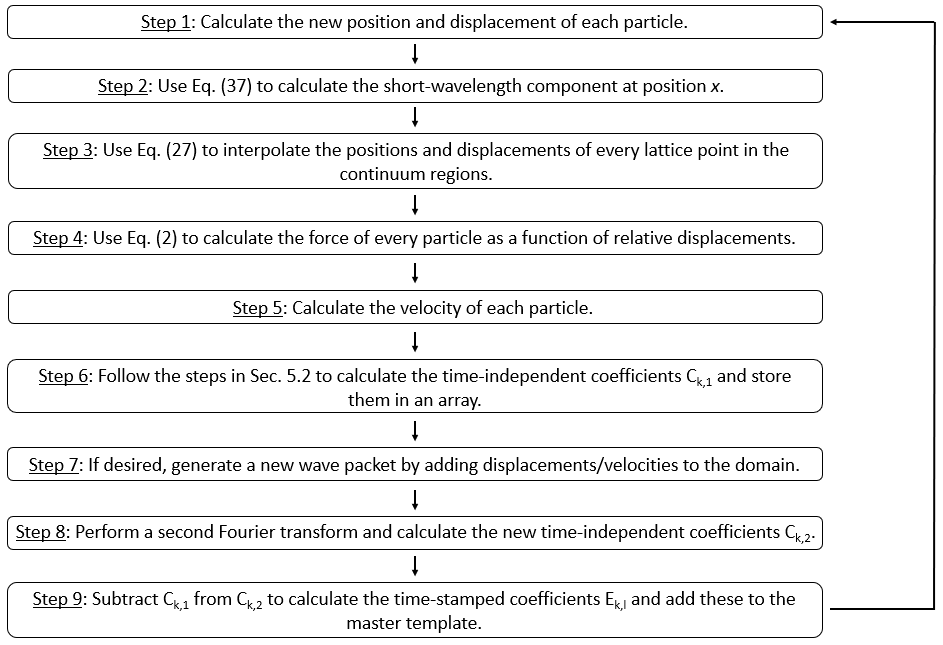}
                \caption{Flow chart showing the various steps taken to pass multiple waves between the atomistic and continuum regions of a CAC domain.}
                \label{Fig:FlowChartPassingWaves}
            \end{figure}

            The flow chart shown in Fig. \ref{Fig:FlowChartPassingWaves} provides an overview of the various steps required to pass more than one phonon wave packet between the atomistic and continuum regions of a CAC domain using the velocity Verlet time integration algorithm.
            We note that the second Fourier transform always occurs at the end of each time step regardless of whether or not a new wave is nucleated.
            If there is not a new phonon present in the domain, the first and second $C_{k,l}$ coefficients will cancel out and $E_{k,l}$ will equal zero.
            As a result, no ``extra" data is ever added to the master template.
    
    \section{Benchmark examples with multiple waves} \label{Sec: Benchmark examples with multiple waves}
\begin{figure}[htpb]
            \centering
            \begin{subfigure}{0.48\textwidth}
                \includegraphics[width=\textwidth]{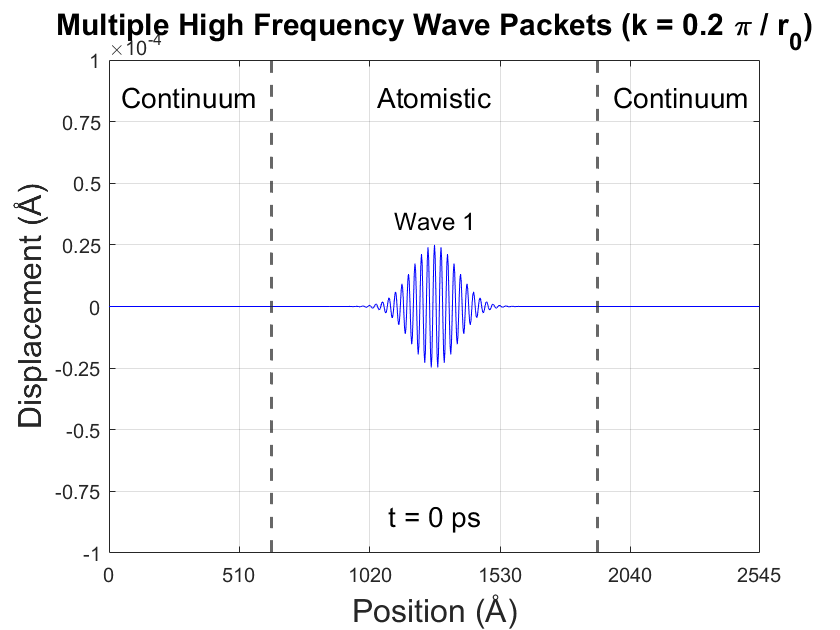}
                \caption{}
            \end{subfigure}
            \begin{subfigure}{0.48\textwidth}
                \includegraphics[width=\textwidth]{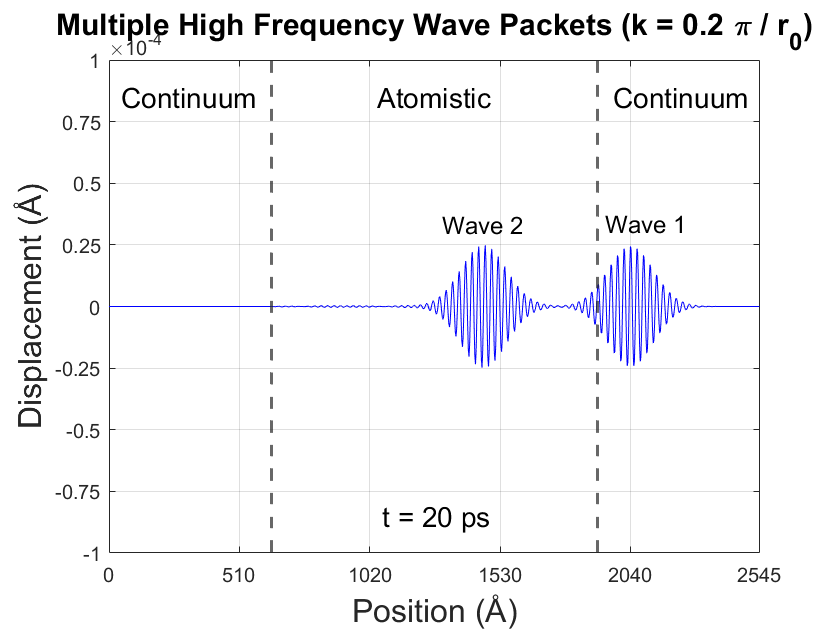}
                \caption{}
            \end{subfigure}
            \\
            \begin{subfigure}{0.48\textwidth}
                \includegraphics[width=\textwidth]{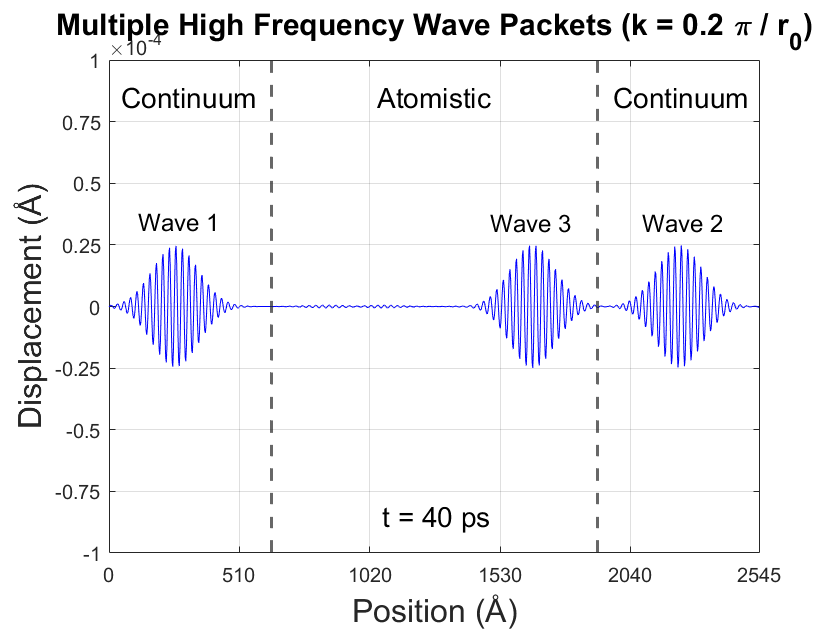}
                \caption{}
            \end{subfigure}
            \begin{subfigure}{0.48\textwidth}
                \includegraphics[width=\textwidth]{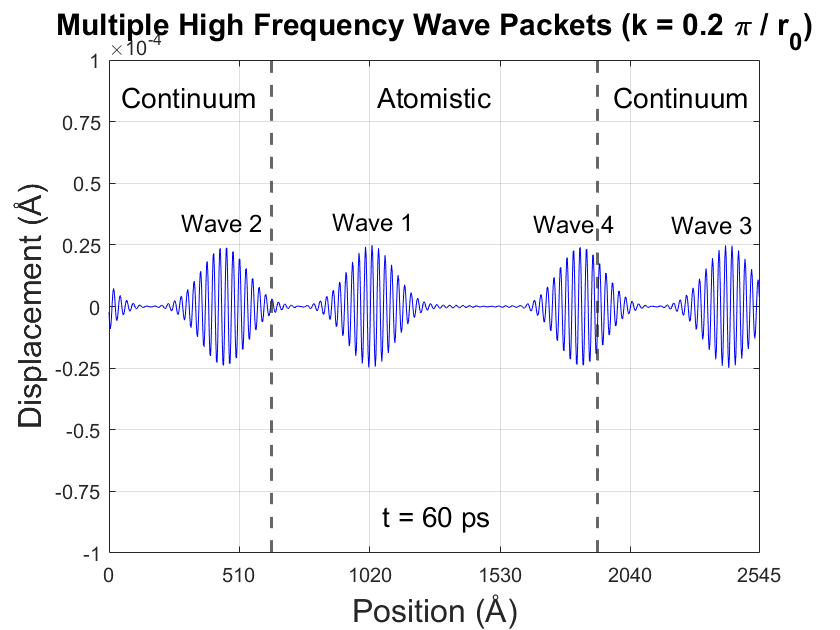}
                \caption{}
            \end{subfigure}
            \caption{Multiple high-frequency phonon wave packets traveling through a single CAC domain. 
            In this case, each phonon has the same wavevector: $k = 0.2 \, \pi/r_0$.}
            \label{Fig:LDMultipleWavePacketSimulation_k2}
        \end{figure}

\begin{figure}[htpb]
            \centering
            \begin{subfigure}{0.48\textwidth}
                \includegraphics[width=\textwidth]{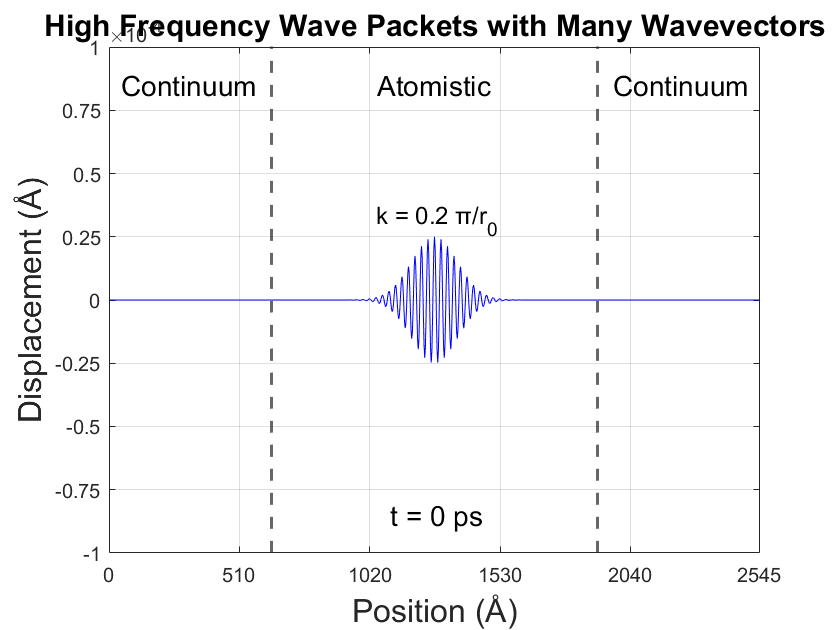}
                \caption{}
            \end{subfigure}
            \begin{subfigure}{0.48\textwidth}
                \includegraphics[width=\textwidth]{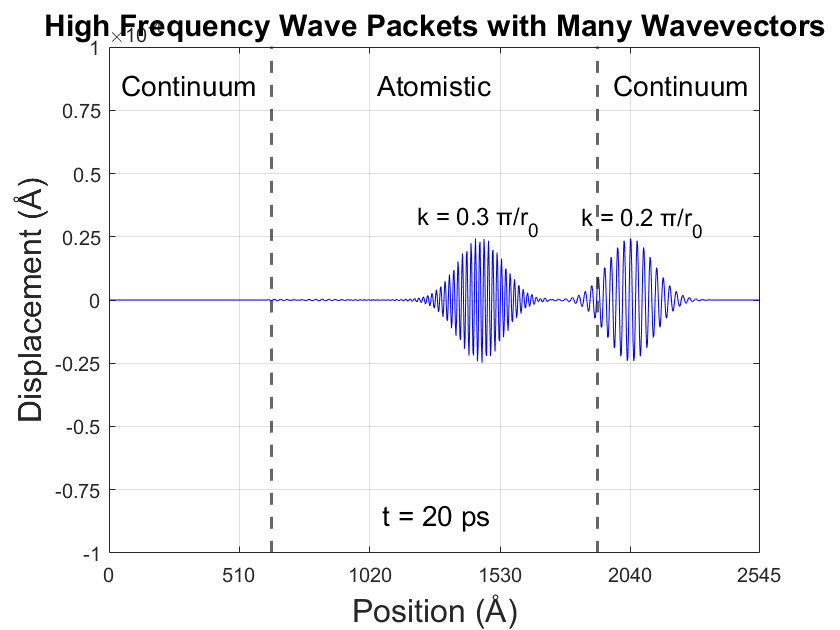}
                \caption{}
            \end{subfigure}
            \\
            \begin{subfigure}{0.48\textwidth}
                \includegraphics[width=\textwidth]{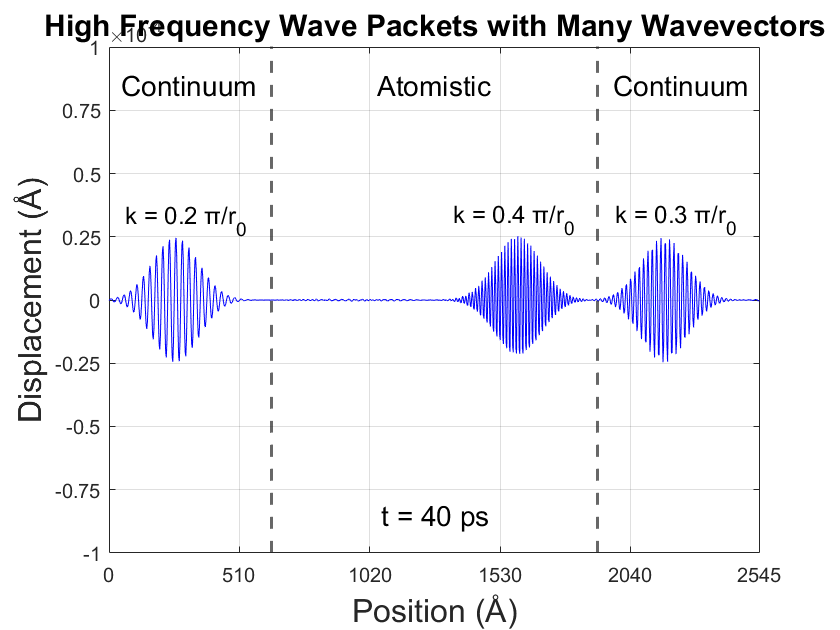}
                \caption{}
            \end{subfigure}
            \begin{subfigure}{0.48\textwidth}
                \includegraphics[width=\textwidth]{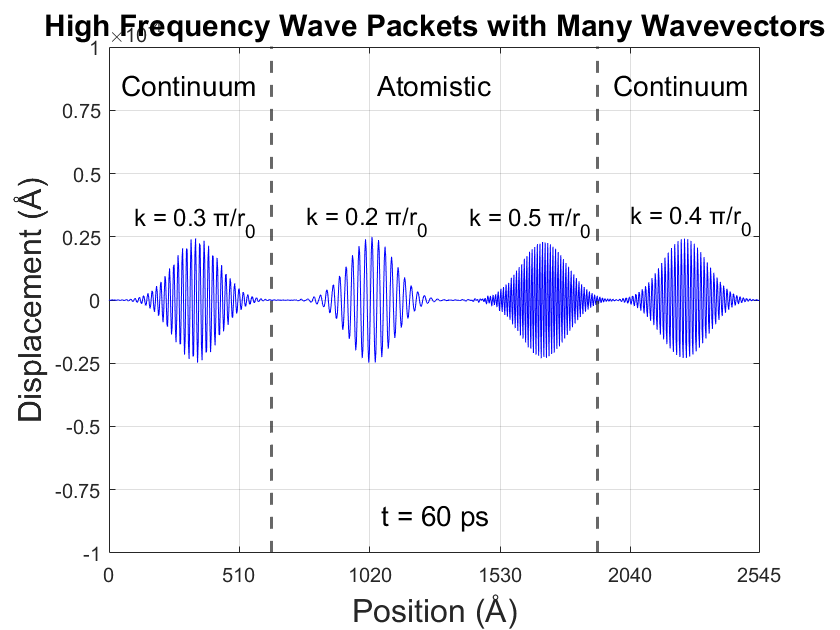}
                \caption{}
            \end{subfigure}
            \caption{Multiple high-frequency phonon wave packets traveling through a single CAC domain. 
            In this case, each phonon has a different wavevector as is shown.}
            \label{Fig:LDMultipleWavePacketSimulation_ManyK}
        \end{figure}

        To verify the implementation and effectiveness of the technique discussed in Sec. \ref{Sec: Upgrading the lattice dynamics technique for multiple waves}, we perform simulations with multiple waves using the CAC framework described in Sec. \ref{Sec: Domain geometry}.
        Specifically, we utilize the new technique to pass various high-frequency wave packets between the atomistic and continuum regions of the multiscale domain. 
        Results from these simulations can be seen in both Fig. \ref{Fig:LDMultipleWavePacketSimulation_k2} as well as in Fig. \ref{Fig:LDMultipleWavePacketSimulation_ManyK}.
        In each simulation, we nucleate four waves in the atomistic region and allow them to propagate to the right and travel across the A-C interfaces. 
        The waves are generated in time increments of $15$ ps, and each has a high wavevector value that would ordinarily cause the phonon to be completely reflected (as demonstrated in Fig. \ref{Fig:WPTMorse}). 
        We note that in Fig. \ref{Fig:LDMultipleWavePacketSimulation_k2}, each phonon has the same wavevector ($k = 0.2 \, \pi/r_0$) while in Fig. \ref{Fig:LDMultipleWavePacketSimulation_ManyK}, the phonons increase in wavevector from $k = 0.2 \, \pi/r_0$ to $k = 0.5 \, \pi/r_0$.
        This is done in order to showcase how the new method can be used with multiple waves of a variety of frequencies within the same domain.
            
        In both figures, we observe that the new method outlined in Sec. \ref{Sec: Upgrading the lattice dynamics technique for multiple waves} permits each short-wavelength phonon wave packet to travel across the A-C interface with no observable reflection.
        Additionally, this scheme facilitates periodic boundary conditions whereby the waves can travel between the two outer continuum regions and back into the inner atomistic region.
        Hence, this method may be used in practical applications which require a periodic domain.
        Finally, we note that this technique can be utilized to track phonons with a variety of frequencies within a single domain, and these waves may interact with each other freely without undermining any stored data. 
        Therefore, we can use this method to transmit many waves across length scales as they contact each other as well as the domain boundary.
        
    \section{Conclusion} \label{Sec: Conclusion}
        In this paper, we developed a technique to transmit multiple high-frequency phonon waves across length scales within a periodic CAC domain.
        Specifically, we utilized the LD interpolation scheme from \cite{chen2018passing} and introduced novel numerical techniques into the framework to update the continuum region with short-wavelength data and track multiple waves across time.
        We first replicated the phonon dispersion relation of the system in order to find the critical wavevector $k_C$ above which the curves of the coarse-scaled region and fine-scaled region diverged.
        Wave packet simulations confirm that phonons with wavevectors $<k_C$ fully transmit across the A-C interface while phonons with wavevectors $>k_C$ completely reflect.
        Next, we described the LD-based finite element scheme developed in \cite{chen2018passing} to transmit a single short-wavelength phonon across length scales.
        A wave packet simulation confirmed the ability of this method to transmit a high-frequency $k = 0.2 \pi/r_0$ phonon from the atomistic to the continuum region with nearly imperceptible reflection.
        
        We then described the technique to pass multiple high-frequency phonons between the atomistic and continuum regions of the CAC framework.
        To implement this method, we first expanded the short-wavelength displacement equation to account for a variety of wave packets nucleated at different time steps.
        However, the coefficients in this equation were still time-independent, and we showcased how this ``naive" approach could only store information for one phonon at a time. 
        Next, we modified the displacement equation to incorporate ``time-stamped" coefficients which encoded the initialization time of each wave.
        During the integration algorithm, we performed two separate Fourier transforms both before and after a new phonon was generated. 
        By obtaining the amplitude coefficients, we effectively took a ``snapshot" of the domain in \textit{\textbf{k}-space} which allowed us to know which information was new.
        The difference in these coefficients was used to calculate the updated time-stamped coefficients which were then stored in a ``master" array.
        Hence, information from multiple phonons was tracked over time, and the displacement field could be updated to incorporate each of these waves into the continuum regions.
        Simulations performed with this technique confirmed its effectiveness in transmitting multiple short-wavelength phonons across the A-C interfaces.
        
        While this technique can be used to transmit multiple short-wavelength wave packets, we note some limitations of this scheme. 
        The framework, in its current state, is incapable of transmitting short-wavelength waves generated due to physical processes such as scattering.
        For example, during impact simulations, a shock wave may interact with a microstructural interface and produce high-frequency transient waves which travel throughout the domain.
        Such a wave will appear in the system during the Verlet integration (Steps 1-5 in Fig. \ref{Fig:FlowChartPassingWaves}).
        However, in the current framework, the short-wavelength wave packet nucleation occurs at a very specific step in the flowchart (Step 8 in Fig. \ref{Fig:FlowChartPassingWaves}) external to the Verlet integration. 
        As such, any high-frequency wave generated during Verlet integration will not be captured and thus not added to the master template.
        We emphasize, however, that the current technique is not meant to be a decisive solution to a complex problem of wave scattering/transmission in multiscale modeling.
        Rather, this method is a step towards tracking a variety of high-frequency waves which are nucleated in a concurrent domain over time.
        
        In the future, we hope to extend this technique to higher-dimensions and use it to study waves in systems with complex microstructures. 
        \added[id=R2,comment={3}]{Such microstructures could arise from particles being randomly oriented within a monatomic lattice or alloyed materials giving rise to intricate particle arrangements within polyatomic crystals.
        For diatomic systems in particular, there would be two branches of the analytical dispersion relation, and in principle, the present formulation could be used to transmit high-frequency optical phonons between the fine-scaled and coarse-scaled regions of the CAC domain and vice versa.}
        We also hope to expand this method to account for different wave types such as elastic waves and orthogonal wavelets.
        \added[id=R2,comment={2}]{Finally, we intend to eventually solve the scattering problem whereby we could transmit across length scales multiple high-frequency waves generated from a physical process such as a moving dislocation or shock impact.}
    
    \section{Acknowledgments} \label{Sec: Acknowledgments}
        This material is based upon work supported by the National Science Foundation under Grant No. $1950488$.
        Financial support was also provided by the U.S. Department of Defense through the National Defense Science and Engineering Graduate (NDSEG) Fellowship Program (F-$1656215698$). 
        Simulations were performed using the Easley computing cluster at Auburn University.
    
    \section{Data availability}
        The raw/processed data required to reproduce these findings cannot be shared at this time due to technical or time limitations.

    \bibliographystyle{ieeetr}
    \bibliography{CAC1D_PassingWaves}
    
    \appendix
    
    \section{Solving for the short-wavelength amplitude in 1D} \label{Sec: Solving for the short-wavelength amplitude in 1D}
        In 1D, there is only one phonon branch ($\nu = 1$), particles can only travel in the $\pm$\textbf{x} direction ($\textbf{e}_{\textbf{k} \nu}^{\alpha} = 1$), and there is only one atom per unit cell ($N_{\alpha} = 1$). 
        As a result, Eqs. (\ref{Eq: Modal amplitude displacements}) and (\ref{Eq: Modal amplitude velocities}) reduce to the following:
        \begin{align}
            C_k &= \sum_{j=0}^{N_A-1} u_j \mathrm{exp} \left[-i \left(k \cdot x_j \right) \right] = A_{k} + B_{k} \label{Eq: Modal amplitude displacements 2} \\
            D_k &= \sum_{j=0}^{N_A-1} v_j \mathrm{exp} \left[-i \left(k \cdot x_j \right) \right] = (B_k - A_k) i \omega_k. \label{Eq: Modal amplitude velocities 2}
        \end{align}
        Solving Eqs. (\ref{Eq: Modal amplitude displacements 2}) and (\ref{Eq: Modal amplitude velocities 2}) for $A_k$ and $B_k$ gives the following:
        \begin{align}
            A_k &= \frac{C_k}{2} + i\frac{D_k}{2 \omega_k} \\
            B_k &= \frac{C_k}{2} - i\frac{D_k}{2 \omega_k}.
        \end{align}
        Substituting these expressions for $A_k$ and $B_k$ back into Eq. (\ref{Eq: Short wavelength amplitude}) when $x_j = x_0$ and $t = 0$, we get the following:
        \begin{align}
            u_s(x_0, 0) &= \sum_k (A_k + B_k) e^{ikx_0} \\
            &= \left[\frac{C_k}{2} + i\frac{D_k}{2\omega_k} + \frac{C_k}{2} - i\frac{D_k}{2\omega_k} \right] e^{ikx_0} \\
            &= C_k e^{ikx_0}.
        \end{align}
        Hence, we arrive at the following expression for the amplitude of the short-wavelength phonon mode in 1D:
        \begin{equation}
            U_k(x_0, t_0) = C_k e^{ikx_0} = \sum_{j=0}^{N_A-1} u_j e^{-ikx_j} e^{ikx_0}.
        \end{equation}
        Substituting this back into Eq. (\ref{Eq: usAlpha_2}) for the one-dimensional monatomic chain:
        \begin{align}
            u_{s}(x, t) &= \frac{1}{N_A} \sum_{k (k > k_C)} C_k e^{ikx_0} \mathrm{exp} \left [i \left(k \cdot \Delta x - \omega_k \Delta t \right) \right] \\
            &= \frac{1}{N_A} \sum_{k (k > k_C)} C_k \mathrm{exp} \left \{i \left[k \cdot (x_0 + \Delta x) - \omega_k \Delta t \right] \right\} \\
            &= \frac{1}{N_A} \sum_{k (k > k_C)} C_k \mathrm{exp} \left [i \left(k \cdot x - \omega_k \Delta t \right) \right] \label{Eq: us_1D_exp}
        \end{align}
        where $x = x_0 + \Delta x$ is the location of the node in the continuum region.
        Additionally, $C_k$ is given by the following expression:
        \begin{equation} \label{Eq: Ck equation exponential}
            C_k = \sum_{j=0}^{N_A-1} u_j e^{-ikx_j} = \sum_{j=0}^{N_A-1} u_j \left[\mathrm{cos}(k \cdot x_j) - i \mathrm{sin}(k \cdot x_j) \right].
        \end{equation}
        As a result, we can rewrite $u_{s}(x, t)$ in trigonometric form as follows:
        \begin{equation} \label{Eq: }
            u_{s}(x, t) = \frac{1}{N_A} \sum_{k (k > k_C)} \left[\mathrm{Re}(C_k) - i \mathrm{Im}(C_k) \right] \left[\mathrm{cos}(k \cdot x - \omega_k \Delta t) + i \mathrm{sin}(k \cdot x - \omega_k \Delta t) \right]
        \end{equation}
        where
        \begin{align} 
            \mathrm{Re}(C_k) = \sum_{j=0}^{N_A-1} u_j \mathrm{cos}(k \cdot x_j) \label{Eq: Ck_real_coefficient} \\
            \mathrm{Im}(C_k) = \sum_{j=0}^{N_A-1} u_j \mathrm{sin}(k \cdot x_j). \label{Eq: Ck_imag_coefficient}
        \end{align}
        Keeping only the real parts, we arrive at our final expression for the short-wavelength displacement in 1D:
        \begin{equation} \label{Eq: ShortWavelengthDisplacement1D}
            u_{s}(x, t) = \frac{1}{N_A} \sum_{k (k > k_C)} \left[\mathrm{Re}(C_k)\mathrm{cos}(k \cdot x - \omega_k \Delta t) + \mathrm{Im}(C_k)\mathrm{sin}(k \cdot x - \omega_k \Delta t) \right].
        \end{equation}
           
        Therefore, when simulating a high-frequency phonon wave packet using the described LD technique, we utilize the velocity-Verlet algorithm from Sec. \ref{Sec: Integration algorithm} to evolve the wave initialized in the atomistic region.
        Next, we store the displacements of each particle in an array at time $t = 0$ ps and follow the procedure outlined in Sec. \ref{Sec: Determining the amplitude of the short-wavelength phonon mode} to calculate $U_{\textbf{k} \nu}^{\alpha}$. 
        Then, at each time step $t$, we use Eq. (\ref{Eq: usAlpha_2}) to compute $\textbf{u}_{s}^{\alpha}(\textbf{x}, t)$ at a given position $\textbf{x}$, and we calculate the total displacement of each continuum node using Eq. (\ref{Eq: LD_Displacement_2}).
        Finally, we calculate the internal force of each particle as a function of relative displacements using Eq. (\ref{Eq: Internal Force Density}) and update the time step.
        This technique allows high-frequency phonons that would ordinarily be reflected at the A-C interface to pass smoothly between the atomistic and continuum regions.
    
\end{document}